\documentclass[preprint,aps,showpacs,floatfix]{revtex4}
\usepackage{epsf}

\begin{document}

\title {Condensation energy in phonon and spin mediated superconductors -
strong coupling approach}  

\author{Robert Haslinger$^1$, and Andrey V. Chubukov$^2$}

\affiliation{$^1$ Los Alamos National Laboratory, Los Alamos, NM 87545}
\affiliation{$^2$ Department of Physics, University of Wisconsin, 
Madison, WI 53706}   
\date{\today}   
 
\begin{abstract}   
We consider the condensation energy, $E_c$ of strongly 
coupled, magnetically-mediated superconductors within
the context of the spin-fermion model.  We argue that
although experimentally obtained values of $E_c$ are of
the same order of magnitude as would be expected from
BCS theory in optimally and overdoped cuprates, 
this agreement is coincidental.  The actual physics
behind the condensation energy is much richer.  In particular,
we argue that it is vital to take both the fermionic and
bosonic contributions to the condensation energy into
account when considering such materials, and that it is only
the sum of the two contributions, $E_c$, which has physical
meaning.  Both the experimental and our theoretically
calculated condensation energies exhibit a {\it decrease}
with further underdoping past optimal doping, e.g. in
the mid to strong coupling regime.  Below optimal doping, the
physics is qualitatively different from BCS theory
 as the gain in the condensation energy 
 is a result of the feedback on 
spin excitations, while the fermionic contribution to $E_c$ is positive
 due to an ``undressing'' feedback on the fermions.
 We argue that the same feedback effect accounts for a gain in  
 the kinetic energy  at strong coupling.
\end{abstract}  
\maketitle

\section{introduction}

 Undertanding the origin of the condensation energy is an 
important step towards identifying the mechanism of 
high temperature superconductivity in the cuprates. 
In a BCS superconductor, the 
condensation energy $E_c$ - the energy gain in a superconductor 
 compared to the normal state at the same $T$, smoothly increases
 below $T_c$ and at $T=0$ reaches 
 $E_c^{BCS} = - V N_f \Delta^2/2$, 
where $V$ is the volume, $\Delta$ is the superconducting gap and 
$N_f = mp_F/(2\pi^2 \hbar^3)$ is the fermionic density of states. \cite{bcs} 
The decrease in the total
energy upon pairing results from a fine 
competition between an increased kinetic energy and a decreased  
potential energy, both of which are much larger than $E_c$. 
The BCS condensation energy  can be experimentally 
extracted from the jump of the specific heat
 at $T_c$ as within the BCS theory $C_s -C_n \approx 6.08 E_c/T_c$.

Since the fermionic density of states  is only weakly dependent on doping, the 
 application of the BCS  formula for the condensation energy  
to cuprate superconductors 
would imply that $E_c$ and $\Delta^2$ scale in the same way.   
However, the measured gap increases monotonically with decreasing 
doping~\cite{arpes}, while the 
jump of the specific heat has a non-monotonic doping dependence.
In the overdoped regime it initally increases with reduced doping,
but below optimal doping further underdoping leads to a
{\it decrease} in the specific heat jump \cite{loram}. 
This discrepancy between the trends in $C_s-C_n$ 
and in $\Delta$ as functions of doping  
clearly makes the applicability of the BCS formula to the cuprates
questionable.

In view of this fact, and the general consensus that
the cuprates are strongly coupled systems, it comes as no surprise
 that effort has already been made to explain
 the cuprate condensation energy as a result of non-BCS physics. 
 Scalapino and White~\cite{scal_white} conjectured that at strong coupling, 
the dominant contribution to the condensation energy comes from a feedback 
effect on the magnetic excitations of the system.
On the other hand,  Norman~{\it et al} 
 argued~\cite{mike1} 
that the condensation energy  likely has an electronic 
 origin, and is driven by a gain in the kinetic energy which at strong
  coupling is negative (in contrast to BCS theory) 
 because of a strong ``undressing'' of fermions which 
  bear a greater resemblance to free particles in the superconducting state 
than they do in the normal state.  Similar ideas were expressed 
 by Hirsh and Marsiglio~\cite{hirsh}. A different idea, related 
 to the lowering of the Coulomb energy in the superconducting state
  has been proposed by 
 Leggett~\cite{leggett}. 

In this communication we argue that these apparently  disparate
viewpoints are in fact consistent with each other, and describe the 
same strong coupling physics.  We argue that at strong coupling, the
relation between $E_c$ and $\Delta$ is qualitatively different from BCS
theory and is consistent with the experimental trends in
the underdoped cuprates.  We show furthermore that the
 strong coupling effects are in large part the result of mutual
feedback between the fermions and bosons.  We make the case that
the contributions to the condensation energy from these two
channels may not be considered independently from each other, and
that only the sum of the two contributions has physical meaning.

Our point of departure is the general  equation
for the free energy of an interacting electron system,
 derived by Luttinger and Ward~\cite{lutt_ward} for the normal state
  and extended to a superconductor by Eliashberg~\cite{eliash}.
We first review strongly coupled phonon superconductors. We discuss the 
 validity of the Eliashberg formula and 
 describe the BCS, Bardeen-Stephen \cite{bardeen_stephen}
  and Wada \cite{Wada}   expressions for the condensation 
energy (for a review, see Ref.~\cite{review}).  The BCS expression 
  neglects both 
 bosonic and fermionic self-energies (apart from a trivial renormalization 
 of the dispersion). The  Bardeen-Steven and Wada expressions include the fermionic 
 self-energy and its change between the normal and the superconducting
  states, but neglect the feedback from superconductivity on phonons.
   The two expressions are generally considered to be identical, 
 however we argue that this is only true as long as the
 feedback from fermions onto phonons is neglected. We review Bardeen and Steven's 
 arguments for the validity of this approximation for phonon superconductors.

  We next modify the Eliashberg
equation for $E_c$ to the case of magnetically
mediated superconductors. We argue that in qualitative distinction
 to the phonon case, the feedback effect from 
 the superconductivity onto the bosons 
 may not be neglected if the pairing is magnetic.   We derive the relevant
equations and explicitly compute $E_{c}$ assuming that the 
 pairing is due to spin-fluctuation exchange and is described 
 by the spin-fermion model. We show that when the feedback on bosons is 
 non-negligible,  
the calculations require care as both electronic and spin parts of the 
condensation energy are ultraviolet divergent. We explicitly show that 
these divergencies are cancelled out between the two terms, and the 
total $E_c$ (which turns out to be the only 
 physically meaningful quantity) is free from divergencies.  We 
 furthermore demonstrate that
 one can avoid the divergencies by performing the computations 
  in real frequencies. 
We apply the results to the cuprates and 
 show that our theoretical $E_c$ agrees
 with the data both in magnitude and in the doping dependence. 
We view this agreement as a 
support for the spin-fluctuation scenario for the cuprates. 

\section{Phonon Superconductors}

 The condensation energy $E_c$ is the difference between the
 free energies in the normal and superconducting states.  
\begin{equation}
E_c =  F_{s} - F_{n}.
\label{1a}
\end{equation}
In the Greens function technique one evaluates the
 grand free energy $\Omega$. The difference $F_n - F_s$ 
 coincides with $\Omega_s - \Omega_n$ provided that the chemical 
 potential $\mu$ does not change between the two states. In the
  Eliashberg theory that we will be using, this is the case as 
  the Fermi energy is assumed to be much larger than $\Delta$, and 
the corrections to $\mu$ due to pairing, 
which scale as powers of $\Delta/E_F$, are
 neglected~\cite{bardeen_stephen}.

There are two ways to compute $E_c$ using Green's functions. 
The first approach is to use the general formula for the ground 
state energy of the interacting fermionic system in terms of 
the integral over the running coupling constant~\cite{mahan}:
\begin{equation}
\Omega = \Omega_0 - i T \sum_k \int_0^\lambda \frac{d \lambda^\prime}{\lambda^\prime} 
G(k,\lambda^\prime) \Sigma (k, \lambda^\prime)
\label{e1}
\end{equation}
where $\Omega_0$ is the Free energy for free fermions, 
$\Sigma$ is the full self-energy, and  $G = 
(i (\omega + \Sigma (k, \lambda^\prime)) - \epsilon_k)^{-1}$ is the full Green's function. 
Both are functions of  a 4 vector of momentum and Matsubara
 frequency $k=({\bf k},\omega_n)$ and the running coupling constant $\lambda^\prime$.
 Note that here and below we define $\Sigma$ with an extra
factor of $i$.

Eq. (\ref{e1}) is applicable to both normal and superconducting states 
(for the latter, $\Sigma (k, \lambda^\prime)$ has a pole), 
 however, is not convenient for our purposes
 as  numerical calculations then require solving the  Eliashberg equations for
 a large set of coupling values. Still, we tried this formalism, and will
 discuss the results in the 
 Appendix A. The approach which we will be using in the bulk of the paper was
 suggested by Luttinger and Ward~\cite{lutt_ward}  who 
demonstrated that  it is possible
to re-express $\Omega$ in the normal state via a series of closed linked skeleton diagrams
with fully dressed fermionic and bosonic
propagators. Their approach was extended to the superconducting state by 
 Eliashberg \cite{eliash}. We refer the reader to Refs.\cite{lutt_ward} and \cite{eliash}
 for the details of the derivation, and here just present the result.
 In the superconducting state, $\Omega$ has the form  
\begin{eqnarray}
\Omega &=& -2T \sum_{k} \lbrace \frac{1}{2} 
\log [\epsilon^2_{\bf k} + \tilde\Sigma^2(k) + \Phi^2(k)] 
- i\Sigma(k)G(k)
+ i\Phi(k)F(k) \rbrace \nonumber\\
&+& \frac{1}{2} T \sum_q \lbrace \log[D^{-1}(q)] +
\Pi(q) D(q) \rbrace \nonumber\\
&+& T^2 \sum_{k,k'} \alpha_{k-k'}^2 \lbrace G(k)D(k-k')G(k') + 
F(k)D(k-k')F(k') \rbrace + \ldots
\label{phononom}
\end{eqnarray}
The last term in the above equation is the sum
of the first two closed linked skeleton diagrams, the dots stand for
 higher order diagrams.
The functions $G(k)$ and $F(k)$ are the real and anomalous
Greens functions given by  
\begin{eqnarray}
G(k) &=&-\frac{\epsilon _{\mathbf{k}}+i
\widetilde{\Sigma}(k)}{\epsilon _{\mathbf{k}}^{2}+\widetilde{\Sigma}
(k)^{2}+\Phi^2(k)},  \nonumber \\
F(k) &=&i\frac{\Phi(k)}{
\epsilon _{\mathbf{k}}^{2}+\widetilde{\Sigma}(k)^{2}+
\Phi^{2}(k)} 
\label{greensfunctions}
\end{eqnarray}
 where  $\Phi(k)$ is the
pairing vertex and    
$\tilde\Sigma(k) = \omega_n + \Sigma(k)$.
The conventionally defined pairing gap 
$\Delta(k)$ is the ratio of 
the anomalous vertex and the self-energy: 
$\Delta(k) = \Phi(k) \omega_m/{\tilde \Sigma}(k)$.
 Finally, $\alpha_k$ is the electron-phonon coupling, and
$D(q)$ is the dressed phonon propagator given by
$D^{-1}(q) = D_0^{-1}(q) - \Pi(q)$ where $\Pi(q)$ is the polarization bubble, 
and $D_0(q)$ is the bare propagator. A careful reader may note that
 we define  $D(q)$ as dimensionless
 quantity (see (\ref{phononom}), while the actual phonon propagator has a dimension of inverse energy.  This does not cause problems, however, as
 $D$ appears only in a combination with 
$\alpha^2$, and the extra  overall energy factor can be absorbed from $\alpha^2_k$.    
 We will be using Eq. \ref{phononom} as the 
point of departure for our analysis.

Eq. \ref{phononom} is quite general.  Its use
for the calculation of $E_c$ takes into
account not just the introduction of the pairing vertex $\Phi$
but also changes in the fermionic self energy $\Sigma$ and 
the polarization bubble $\Pi$.  Thus it calculates
the contributions to $E_c$ from both the fermions {\it and} bosons.
 
The approximations made in the phonon case are associated with the
 smallness of the sound velocity $v_s$ compared to the Fermi velocity $v_F$.
It turns out that higher order closed linked diagrams 
 form series in powers of $\lambda v_s/v_F$, where $\lambda$ is the
 dimensional coupling constant which scales with $\alpha_{k-k^\prime}$ (see below).
  The approximation, attributed to Migdal~\cite{migdal} 
 and Eliashberg~\cite{eliash1}, is to neglect $O(\lambda v_s/v_F)$ terms
  without assuming that $\lambda$ is by itself small. This would imply 
 neglecting all terms labled as dots in (\ref{phononom}). 

 The physics behind the Migdal-Eliashberg approximation is best revealed by 
analyzing the  perturbation series for $\Sigma(k)$ 
and $\Phi (k)$. These series can be separated  into
 terms  that scale as powers of 
$\lambda$ and  terms that scale as powers of $\lambda v_s/v_F$. 
The perturbative series in powers of $\lambda v_s/v_F$ 
arises from the actual 
electron-phonon scattering, and  the factor $v_s/v_F$ results from  
 the fact that in this process 
electrons are forced to vibrate at phonon frequencies 
 far from their own resonance. This gives rise to vertex 
 corrections (at typical pairing frequencies), and to
 an equal 
  renormalization of fermionic $\omega$ and $\epsilon_k$, i.e.
 the quasiparticle residue is renormalized, but the quasiparticle 
 mass  remains a bare one. The vertex correction
 diagrams for $\Sigma (k)$ and $\Phi (k)$, embedded into the closed linked
diagrams, gives rise to higher-order skeleton diagrams for the Free energy,
and it is these diagrams which are dropped.

 The  terms that form series in $\lambda$ 
 are different and they are not considered to be small.
 In the normal state, they can be understood as coming from phonon-induced  
interactions between electrons and their own zero-sound collective modes. 
These terms do not contribute to the vertex renormalization (at typical
 frequencies for the pairing), and gives rise to a $\Sigma(k)$ that only 
 depends on frequency. The $O(\lambda)$ terms also contribute to the 
 pairing problem and in the
 superconducting state give rise to $\Phi (k)$ that again depends only 
 on frequency. 

Neglecting the higher order skeleton diagrams, Eliashberg obtained
 the closed-form expression for $\Omega_{s}$ given in Eq. \ref{phononom}.
  The closed form, coupled
 equations  for the fermionic
 self-energy $\Sigma (k) = \Sigma (\omega)$ and the pairing vertex 
$\Phi (k) = \Phi (\omega)$ then 
follow from the condition that $\Omega_s$ given by (\ref{phononom})
 is stationary with respect to variations in $\Sigma (k)$ and $\Phi (k)$. 
The conditions $\delta \Omega_{s}/\delta \Sigma (k) =
 \delta \Omega_{s}/\delta \Phi (k) =0$ yield
\begin{eqnarray}
{\tilde \Sigma}(k) &=&  \omega + i T\sum_{k'} \alpha^2_{k-k'}G(k')D(k-k') \nonumber\\
\Phi(k) &=&  - i T\sum_{k'} \alpha^2_{k-k'} F(k')D(k-k')
\label{2ndord}
\end{eqnarray}

The Migdal-Eliashberg approximation has two further implications. 
First, the fermionic dispersion may be approximated  
by $\epsilon_k = v_F (k-k_F)$, as typical pairing frequencies 
 should be much smaller than $E_F$ if 
 $\lambda v_S/v_F$ is to be small.  Second, the momentum integration 
 over $k^\prime$ can be factorized: the integration
 over momenta  transverse to the Fermi surface involves only 
normal and anomalous fermionic propagators, via $\epsilon_k$, while 
the integration along the Fermi surface involves only the bosonic propagator 
 in which one can set $|k| = |k'| =k_F$, i.e., deviations 
 from the Fermi surface are neglected leading to a bosonic 
 propagator dependant only on frequency.  Corrections to this approximation
 again scale as $\lambda v_s/v_F$.
Finally,  it is assumed (without justification) 
that the Fermi surface is isotropic in the sense that
 $v_F$ is considered to be independent of $k$.
Under these approximations, the momentum integration in 
the equations for $\Sigma (\omega)$ and $\Phi (\omega)$
can be performed exactly.
Approximating the momentum sum by an integral over energy and 
 an associated density of states $N_f$ as 
\begin{equation}
\sum_k \rightarrow N_f \int_{-\infty}^{\infty} d\epsilon
\end{equation}
, inserting equations \ref{greensfunctions} into equation
\ref{2ndord} and performing the integration we obtain 
\begin{eqnarray}
{\tilde \Sigma} (\omega) &=& \omega  + 
 \pi T N_f \sum_{\omega'} \alpha^2_{\omega-\omega'}
D(\omega -\omega') \frac{{\tilde 
\Sigma} (\omega')}{\sqrt{\Delta^2 (\omega') + {\tilde \Sigma}^2 (\omega')}} 
 \nonumber\\
\Phi(\omega) &=&  \pi T N_f 
\sum_{\omega'} \alpha^2_{\omega - \omega'} D(\omega - \omega')  
\frac{\Phi (\omega')}{\sqrt{\Delta^2 (\omega') + {\tilde \Sigma}^2 (\omega')}} 
\label{2ndord_1}
\end{eqnarray}

An essential feature of the above equations is that apart from
the assumption of the smallness of $\lambda v_s/v_F$, it is assumed
assumed that the phonon polarization bubble $\Pi (k)$, which accounts for the 
effects of the electrons on phonons,
  may  be neglected.
 Analagously to the derivation of equation \ref{2ndord}, an expression
 for $\Pi (k)$ can be formally obtained from
 the feature that $\Omega_s$ given by (\ref{phononom}) 
is also stationary with respect to variations in $\Pi (k)$. The condition
$\delta \Omega_s/\delta \Pi (k) =0$ yields
\begin{equation}
\Pi(k)  = -2 T D(0) \sum_{k'} \alpha^2_{k-k'}
[G(k)G(k-k') + F(k)F(k-k') ]
\end{equation} 
The fact that $\Pi (k)$ is irrelevant to the phonon problem
is not immediately apparent and this issue must
be considered carefully.
In the normal state, 
\begin{equation}
\Pi (k) \propto \lambda \frac{v_s}{v_F}~\frac{\omega_m}{\omega_D}~  
\frac{p_F}{|{\bf k}|}
\end{equation}
 reflecting the decay of a low-energy bosonic 
 mode into a fermionic particle-hole pair. 
This decay term obviously cannot be neglected at the lowest frequencies as it
 accounts for the leading low-frequency dependence of $D (\omega_m)$. I.e. for 
an Einstein phonon, 
$D_0 (\omega_m) = 2\omega^2_D/(\omega^2_D + \omega_m^2)$
 where $\omega_D$ is the Debye frequency (we recall that $D (\omega_m)$ 
 is dimensionless, hence 
$D^{-1} (\omega_m) = D^{-1}_0 (\omega_m) - \Pi (k)
 \approx 1 - \Pi (k)$ at the lowest frequencies.
However, for $\lambda = O(1)$, 
the frequencies relevant to the pairing problem are of order  
$\omega_D$, typical momenta are of order $p_F$, 
and hence $\Pi (k)$ at typical frequencies 
is small to the same extent as 
$\lambda v_s/v_F$ and other similar terms in the Eliashberg theory.   
In the superconducting state, the low-energy phonons are gapped, and
 at frequencies smaller than $\Delta$,
\begin{equation}
\Pi (k)  \propto 
\lambda \frac{v_s}{v_F}~\frac{\omega_m}{\Delta}~\frac{\omega_m}{\omega_D}~  
\frac{p_F}{|{\bf k}|}.
\end{equation} 
This difference
between $\Pi (k)$ in the normal and superconducting states reflects
a fundamental change in the bosonic dynamics at small frequencies
 due to the gapping of low-energy fermionic excitations. 
Still, however, at typical frequencies for the pairing, 
$\omega \sim \Delta \sim \omega_D$, $\Pi(k) \sim \lambda v_s/v_F \ll 1$, 
and the polarization 
operator can be neglected compared to the bare phonon propagator. 
This was shown explicitly by
 Bardeen and Stephen~\cite{bardeen_stephen} who evaluated the 
 contribution to $E_c$ from 
  the change of the 
 bosonic dynamics between normal and superconducting states and 
demonstrated  
 that in the phonon problem the bosonic piece  in $E_c$ is small 
compared to the electronic piece and hence can be safely neglected.

We will see below that neglecting the bosonic contribution to $E_c$ 
is {\it not} a justifiable approximation if one is dealing with magnetic
 superconductors
where the bosonic mode is a collective mode of the fermions.  This
is the main difference between phonon and magnetic superconductors, 
and we will examine this issue in the next section. 
 
For completeness, we present  several 
useful forms for $\Omega_S$ and $E_c$ for the phonon case. 
For simplicity, we assume that the 
 electron-phonon coupling is independent of frequency, and the 
phonon spectrum consists of a single Einstein boson with a frequency 
$\omega_D \sim v_s p_F$, i.e., 
\begin{eqnarray}
\alpha_{\omega} &=& \alpha \nonumber\\
D(q) &=& D(\omega_m) = \frac{2\omega^2_D}{\omega_m^2 + \omega_D^2}
\end{eqnarray}

Integrating over momentum in Eq. (\ref{phononom}) as before, we obtain 
\begin{eqnarray}
\Omega &=& -N_f \lbrace 2\pi T \sum_m \frac{\omega_m \tilde\Sigma_{\omega_m}}
{\sqrt{\tilde\Sigma_{\omega_m}^2 + \Phi_{\omega_m}^2}} \nonumber\\
&+& T^2 \pi^2{\bar \alpha}^2 \sum_{m,m'} \frac{\tilde\Sigma_{\omega_m}\tilde\Sigma_{\omega_{m'}}
+ \Phi_{\omega_m}\Phi_{\omega_{m'}}}
{\sqrt{\tilde\Sigma_{\omega_m}^2 +\Phi_{\omega_m}^2}
\sqrt{\tilde\Sigma_{\omega_{m'}}^2 +\Phi_{\omega_{m'}}^2}}
\frac{1}{(\omega_m -\omega_{m'})^2 + \omega_D^2} \rbrace
\label{phononsoln1}
\end{eqnarray}
where we introduced  $\bar\alpha^2 = 2 \alpha^2 N_f \omega^2_D$.
The dimensionless coupling $\lambda$ introduced above is related to 
${\bar \alpha}^2$ as 
\begin{equation}
\lambda = \frac{{\bar \alpha}^2}{\omega^2_D} 
\end{equation}

Eq. (\ref{phononsoln1})  may be simplified by making the standard
substitutions $\tilde\Sigma_{\omega_m} = \omega_m + \Sigma_{\omega_m} =
\omega_m Z_{\omega_m}$ and $\Phi_{\omega_m} = \Delta_{\omega_m} \tilde\Sigma_{\omega_m}/
\omega_m = \Delta_{\omega_m} Z_{\omega_m}$. Substituting these forms into 
(\ref{phononsoln1})  we obtain 
\begin{eqnarray}
\Omega &=& -N_f \lbrace 2\pi T \sum_m \frac{\omega_m^2}
{\sqrt{{\omega_m}^2 + \Delta_{\omega_m}^2}} \nonumber\\
&+& T^2 \pi^2{\bar \alpha}^2 \sum_{m,m'} \frac{{\omega_m}{\omega_{m'}}
+ \Delta_{\omega_m}\Delta_{\omega_{m'}}}
{\sqrt{{\omega_m}^2 +\Delta_{\omega_m}^2}
\sqrt{{\omega_{m'}}^2 +\Delta_{\omega_{m'}}^2}}
\frac{1}{(\omega_m -\omega_{m'})^2 + \omega_D^2} \rbrace
\label{phononsoln2}
\end{eqnarray}
Thus we see that the free energy for a phonon superconductor
is dependent only upon the form of
the gap $\Delta_{\omega_m}$ and is not explicitly upon
the self energy $\Sigma_{\omega_m}$.

A comment is in order here. The Luttinger-Ward  
result for the Free 
energy as a series of closed linked skeleton diagrams  
is, strictly speaking, only valid  for the minimum of $\Omega$,
 i.e., for 
the self-energy that
 satisfies the stationary condition. Otherwise, the Luttinger-Ward generating
 functional (which is what they actually calculated) does not necessarily
  coincide with  the Free energy.
In the normal state, this does not cause a problem with Eq. 
(\ref{phononsoln2}) as $\Omega_N$ does not explicitly depend on $\Sigma_N$. 
(We use the capital subscripts "S" and "N" to denote "superconducting state"
and "normal state" respectively.  Lower case "n" (and "m") refer to
Matsubara frequencies.)
In the superconducting state, Eq (\ref{phononsoln2})  does not imply that 
$\Omega_S$ is at a minimum, i.e., $\delta \Omega_S/\delta \Delta =0$.
 We didn't analyze in detail the corrections to Eq. (\ref{phononsoln2})
which would stem from the difference between the Luttinger-Ward-Eliashberg 
 functional and the actual Free energy, but the estimates show that these corrections would be again be small in $\lambda v_s/v_F$. If this is the case, then
 Eq. \ref{phononsoln2} can be used
 for the study of the profile of the
free energy, i.e. how it evolves for different solutions of the gap. 
In particular, we verified that for the BCS case, 
the expansion of Eq.  \ref{phononsoln2} near $\Delta =0$ to order $\Delta^2$ 
yields the sign  change of the slope at exactly the BCS transition temperature 
$T_c$.  

For the remainder of this communication, we will only be considering
$\Omega$ evaluated
 at the equilibrium solution of $\Delta_{\omega_m}$ when the applicability
 of the Luttinger-Ward formalism
 is rigorously justified.  
To this end, the Eliashberg equation for the equilibrium solution
 $\Delta_{\omega_m}$  may be obtained
by minimizing equation \ref{phononsoln2} with respect to $\Delta$. This yields
\begin{eqnarray}
\Delta_{\omega_m} = \pi T \bar\alpha^2 \sum_{m'} \frac{1}{(\omega_m - \omega_{m'})^2
+ \omega_D^2} \frac{1}{\sqrt{\omega_{m'}^2 + \Delta_{\omega_{m'}}^2}}
\lbrace \Delta_{\omega_{m'}} - \frac{\omega_{m'}}{\omega_m} \Delta_{\omega_m} \rbrace
\label{delsoln}
\end{eqnarray}
This solution for $\Delta$  must then be substituted into
equation \ref{phononsoln2} allowing the calculation of the free 
energy at its minimum, and hence $E_c=\Omega_S - \Omega_N$
where $\Omega_N$ is obtained from $\Omega_S$ by setting $\Delta=0$.
Performing the computations, we obtained Eq. (\ref{wada}) below.

There is, however, a simpler  way to proceed 
towards $E_c$.  In this approach, one specifies at the
outset that one
is only interested in the Free energy at equlibrium. In this
case, the fact that equation \ref{phononom} should be stationary with
respect to variations in $\Sigma$ and $\Phi$ can be invoked 
even before the momentum integration is performed~~\cite{bardeen_stephen,Wada,review}.
Substituting equation (\ref{2ndord}) into equation \ref{phononom}
  and droping the phonon piece, we obtain 
\begin{eqnarray}
\Omega &=& -T \sum_{p} \lbrace  
\log[ \epsilon^2_{\bf k} + \tilde\Sigma^2_{\omega_m} + \Phi^2_{\omega_m}] 
- i\Sigma_{\omega_m}G(k)
+ i\Phi_{\omega_m}F(k) \rbrace 
\label{aaa4}
\end{eqnarray}
Introducing a density of states $N_f$, integrating over $\epsilon_k$  and
subtracting the normal state result from the superconducting result
gives an expression for the condensation energy
first derived by Wada~\cite{Wada}.
\begin{eqnarray} 
E_c &=& - N_f \pi T \sum_m (\sqrt{\widetilde{
\Sigma}^2_{S,\omega_m}+\Phi^2_{\omega_m}} - |{\widetilde \Sigma}_{N,\omega_m}|
 \nonumber\\
 &+& |\omega_m| \frac{|{\widetilde \Sigma}_{S,\omega_m}| - \sqrt{\widetilde{
\Sigma}^2_{S,\omega_m}+\Phi^2_{\omega_m}}}{\sqrt{\widetilde{
\Sigma}^2_{S,\omega_m}+\Phi^2_{\omega_m}}})
\label{wada}
\end{eqnarray}
Let us clearly state what the Wada expression calculates.
It is the strong coupling result for the condensation energy
at thermodynamic equlibrium, under the assumption that there
are no appreciable changes in the bosonic mode between
the normal and superconducting states.  In other words, it
accounts for the appearance of the pairing vertex, as well as
any changes to the fermionic self energy, but
ignores any feedback effects between bosons and fermions.

An equivalent expression for $E_c$,  more  advantageous for 
numerical calculations due to a faster convergence at high frequencies
 was obtained by Bardeen and Stephen~\cite{bardeen_stephen}.
They noticed that there exists an 
 integral relation between $\Sigma_{N, \omega}$ and $\Sigma_{S,\omega}$
\begin{equation} 
 N_f \pi T \sum_m \Sigma_{N, \omega} \frac{|{\widetilde 
 \Sigma}_{S,\omega_m}|}{\sqrt{\widetilde{
\Sigma}^2_{S,\omega_m}+\Phi^2_{\omega_m}}} - \Sigma_{S,\omega} =0
\label{relation}
\end{equation}
that in turn is the consequence of the fact 
 that 
\begin{equation}
T \sum _m \int d^2 k \Sigma_{N,\omega_m} G_{S, \omega_m} (k) =
T \sum _m \int d^2k \Sigma_{S,\omega_m} G_{N, \omega_m} (k)
\label{equality}
\end{equation}
as both quantities can be re-expressed as 
a cross-product  
\begin{equation}
T^2 \sum_{m,n} \int d^2 k D_{\omega_n-\omega_m} G_{N,\omega_m} (k)
G_{S,\omega_n} (k)
\label{expand}
\end{equation}
Using equation (\ref{relation}), Bardeen and Stephen obtained 
 the following expression for
 $E_{c,el} = E_c$. 
\begin{equation}
E_{c,el} = - N_f \pi T \sum_m 
\left(\sqrt{\widetilde{
\Sigma}^2_{S,\omega_m}+\Phi^2_{\omega_m}} - 
|{\widetilde \Sigma}_{S,\omega_m}|\right)~\left(1 - 
\frac{|{\widetilde \Sigma}_{N,\omega_m}|}{\sqrt{\widetilde{
\Sigma}^2_{S,\omega_m}+\Phi^2_{\omega_m}}}\right)
\label{bardeen}
\end{equation}
The practical importance of Bardeen-Stephen result 
 is that at high frequencies, when ${\widetilde \Sigma}_{S,\omega_m} \approx 
{\widetilde \Sigma}_{N,\omega_m} \approx \omega_{m}$, 
 the integrand in (\ref{bardeen}) behaves as $\Phi^4_{\omega_m}/\omega_m^3$, 
 and  the frequency summation rapidly converges. This makes 
 the Bardeen-Stephen expression  more convenient for numerical
 computations than Wada's expression. 

 We emphasize again that the equivalence of the two forms 
for $E_c$ is the consequence of the fact that in the phonon case, 
the change of the bosonic self-energy $\Pi (\omega)$ between 
normal and superconducting states can be neglected, and the accuracy of this
 approximation is governed by the same parameter $\lambda v_s/v_F$ as 
 the accuracy of the Eliashberg theory. In the next section we show 
 that this is {\it not} the case for spin mediated pairing.
We will see  that for superconductors
with an electronic pairing mechanism, the feedback on the pairing boson
 plays a crucial role, and  the Wada's expression for $E_c$
 would give competely
erroneous results. Instead, the full expression \ref{phononom}
must be used.

We pause now to connect with the BCS result for the condensation energy.  BCS  is a 
weak coupling theory.  It assumes that the only change
between the normal and superconducting states is the introduction
of the pairing vertex $\Phi$ which is given by the BCS gap $\Delta$.
 ($\Phi_{BCS}=\Delta$). The fermionic and bosonic
 self energies are both  taken to be negligable.  
The BCS condensation energy is therefore calculated from the
Wada equation with the substitution
 $\tilde\Sigma_{\omega_m} = \omega_m + \Sigma_{\omega_m} \approx \omega_m$.
 Taking
the zero temperature limit we obtain after some simple algebra
\begin{eqnarray}
E_c^{BCS} = -N_f\int_0^\infty \frac{2\omega^2 + \Delta^2}
{\sqrt{\omega^2 + \Delta^2}} -2\omega = -N_f\frac{\Delta^2}{2}
\label{bcs1}
\end{eqnarray}
This is the result that we already cited in the Introduction.
We also see that the frequency integration in (\ref{bcs1}) is 
 confined to $\omega \sim \Delta$, i.e., the condensation energy 
 comes from fermions in a narrow region around the Fermi surface.  
Although this result looks rather straightforward, the 
issue of which fermions contribute to the condensation energy 
 in the BCS case is nontrivial, and we discuss it in detail in Appendix 
B.
 
\section{Magnetic Superconductors}

We now proceed to the case of {\it magnetically} mediated pairing.
The bosonic mode that mediates the pairing is now the low-energy 
 spin susceptibility. 
The Luttinger-Ward formalism, which deals with an arbitrary bosonic mode
 is still valid, i.e., the Free energy has the same form as in 
Eq. \ref{phononom}. The only immediate modification is that now 
the second term of Eq. \ref{phononom}
has an extra factor of 3 reflecting the fact that all three components 
of the spin susceptibility contribute equally to the pairing.  
The Free energy then has the form   
\begin{eqnarray}
\Omega &=& -2T \sum_{p} \lbrace \frac{1}{2} 
\log [ \epsilon^2_{\bf k} + \tilde\Sigma^2(k) + \Phi^2(k)] 
- i\Sigma(k)G(k)
+ i\Phi(k)F(k) \rbrace \nonumber\\
&+& \frac{3}{2} T \sum_q \lbrace \log[{D^{-1} (q)}] +
\Pi(q) D(q) \rbrace \nonumber\\
&+& T^2 \sum_{k,k'} g_{k-k'}^2 \lbrace G(k)\chi(k-k')G(k') 
+ F(k)\chi(k-k')F(k') 
\rbrace + \ldots
\label{spinom}
\end{eqnarray}
The dimensionless bosonic propagator $D(q)$ is now related 
 to the magnetic susceptibility
$\chi_{ij}(q) = \chi (q) \delta_{ij}$ as $D(q) = \chi (q)/\chi 
({\bf Q}, 0)$ where ${\bf Q}$ is the momentum at which the static
 susceptibility is peaked. 
 Similarly to phonons, $\chi (q)$ is  
related to the bare susceptibility by
$\chi^{-1}(q) = \chi_0^{-1}(q) - \Pi(q)$.  
We discuss the exact form of $\chi(q)$ below.
Finally, $g_q$ is now the coupling between the fermionic propagator
and bosonic (magnetic) mode.

At this point the formalism is quite general, taking into account 
all changes to both the fermions and bosons.
In order to proceed further, we will need to assume a specific
model that will allow us to neglect higher order terms in 
Eq. \ref{spinom}. 
 We choose the spin-fermion model in
which fermions are paired via their own collective spin excitations.
Several authors have demonstrated that 
 the exchange of collective
 spin fluctuations peaked at or near the antiferromagnetic momentum
  $Q = (\pi, \pi)$ yields an attraction in the $d_{x^2 -y^2}$ pairing channel. 
We will be studing $E_c$ for this kind of pairing.  

The spin-fermion model is described by the effective action  
\begin{eqnarray}
S &=&-\int_0^\beta d\tau \int_0^\beta d\tau' \sum_{{\bf k},\sigma}
c^\dagger_{{\bf k}\sigma} (\tau)
G^{-1}_0 ( {\bf k}, \tau-\tau') c_{{\bf k}\sigma}(\tau')       \nonumber \\
&& +  \frac{1}{2} \int_0^\beta d\tau \int_0^\beta d\tau'  
\sum_{{\bf q}} \chi_0^{-1} ({\bf q}, \tau-\tau') \, {\bf S}_{\bf q}(\tau) 
\cdot {\bf S}_{-{\bf q}} (\tau')\,                               \nonumber \\
&&  + g_q \int_0^\beta d\tau \sum_{{\bf q}}   \, 
{\bf s}_{\bf q}(\tau) \cdot {\bf S}_{-{\bf q}} (\tau)\, ,        \label{sfm}
\end{eqnarray}
where $G^{-1}_0 ( {\bf k}, \tau) = \partial_{\tau} - 
 {\bf v}_k ({\bf k}-{\bf k}_F)$
 is the bare Fermionic propagator, 
$c^{\dagger}_{{\bf k}, \alpha} $ is the fermionic creation operator
for an electron with crystal momentum ${\bf k}$ and spin projection $\alpha$,
${\bf s} = c^\dagger {\bf \sigma} c$, where 
 $\sigma_i$ are the Pauli matrices,  and 
 $g$  is the coupling constant which measures the strength of the 
interaction between fermionic spins and the collective spin
degrees of freedom described by  bosonic variables $S_q$. For 
simplicity, below we assume  $g_q = g$ to be momentum independent. 
The bare spin susceptibility $\chi_0^{-1} ({\bf q}, \tau)$ 
is assumed to be peaked at $Q$ and has a standard 
Ornstein-Zernike form, i.e., its Fourier transform over $\tau$ is
\begin{equation}
\chi_0 ({\bf q, \omega}) = \frac{\chi (Q)}{1 + 
 \xi^2 ({\bf q}-{\bf Q})^2 - (\omega/(v_s \xi^{-1})^2}\, .             
\label{chi0}
\end{equation}
where $\xi$ is the magnetic correlation length.  
This bare susceptibility comes from fermions with energies
 comparable to $E_F$ 
and should be considered as  an 
 input for the low-energy theory. 

The dimensionless coupling constant for 
the model of Eq. (\ref{sfm}) (defined such that $\Sigma (\omega, k_F)
 = \lambda \omega$  in $D=2$) is, 
\begin{equation} 
\lambda = 4{\bar \omega}/(3 v_F \xi^{-1}),
\end{equation}
where ${\bar \omega} =  9 g^2 \chi(Q)/(16 \pi \xi^2)$. 
 The numerical factors are choosen for further convenience. 
This overall scale ${\bar \omega}$ and the coupling $\lambda$
 are the only two  parameters that matter 
at strong coupling. Other parameters, e.g., $v_s$, turn out to 
be irrelevant (see below). Note that ${\bar \omega}$ is in fact 
independent of $\xi$ as $\chi(Q)$ by itself scales as $\xi^2$. 
 
Near a magnetic transition, $\xi$ is large, i.e., $\lambda \geq 1$, and 
 spin-fermion model is a strong coupling theory in which
feedback effects between fermions and bosons are extremely important.
It has been discussed in depth in \cite{abanov_advances} as a theoretical
 model as well as with respect to the cuprates, and provided an
 explaination of many unusual properties of the cuprates such as 
 non-Fermi liquid behavior in the normal state \cite{robarpes},
 $d_{x^2-y^2}$ pairing \cite{d}, 
and the pseudogap \cite{artem}. 
We note, though, that there is still a great deal
 of contraversy regarding the full description of these phenomena.
Since, however, it is generally believed that the
cuprates are strongly coupled superconductors,
it is instructive, regardless of one's prejudices, 
to examine the condensation energy for the spin-fermion 
model in detail, so as to illustrate the importance
of properly accounting for all feedback effects when dealing
with a strongly coupled system.  We will show that the spin-fermion
model accounts for many aspects of the experimentally measured
condensation energy.

\subsection{The validity of the Eliashberg approximation}

We begin by briefly 
discussing the validity of the Eliashberg approximation
for anti-ferromagnetically mediated superconductivity and how
the assumptions inherent in the Luttinger-Ward condensation energy
formalism are justified. 
For the purposes of calculating the condensation energy,
there are two main issues to be discussed.  First, as with phonons,
it is possible in the spin-fermion model to
separate the perturbative series such 
that the terms resulting from
vertex corrections and higher order diagrams are "small" and
therefore irrelevant.  Second, since the effects of pairing
are greatest near "hot spots" (points on the Fermi surface
connected by ${\bf Q}=(\pi,\pi)$) one can invoke an
effective {\it momentum independence} for the problem, while still
retaining the $d_{x^2-y^2}$ pairing symmetry.
We briefly emumerate the reasoning leading to the above conclusions
here.  A more detailed discussion
 can be found in Ref.\cite{abanov_advances}. 

\begin{enumerate}
\item
Spin fluctuations are collective modes of fermions, hence 
there is no difference between the Fermi velocity and the spin
 velocity, i.e., $v_s \sim v_F$. Then $\lambda v_s/v_F \sim \lambda$,
  i.e., there is no way to separate perturbative series based on
   the difference between velocities. From this perspective, 
   there is no Migdal theorem 
for spin fluctuations, and the perturbation theory with the {\it bare} spin 
 propagator just holds in powers of the coupling $\lambda$. 
\item
The absence of small $v_s/v_F$ implies in turn that the polarization 
 operator $\Pi (Q, \omega_m) = \Pi_{\omega_m}$ is not negligible, as 
 it is for phonons, but is rather dominant
 for $\lambda \gg 1$. The consequence of this is that 
one must simultaneously solve for both the fermionic and bosonic
 self-energies.
\item
In the normal state, $\Pi_{\omega_m} \propto \omega_m$ 
at low frequencies, i.e.,  when $\Pi_{\omega_m}$ dominates the frequency 
 dependence of the spin susceptibility (strong coupling), spin fluctuations   
 become diffusive. This transmutation of the spin dynamics 
 from propagating with $v_s \sim v_F$ for $\lambda \ll 1$ to diffusive 
 for $\lambda \geq 1$ implies that at strong coupling
 bosons become soft compared to electrons. Such softness of bosons, 
  is precisely the physics behind the Migdal theorem. 
 Not surprisingly then, the diagramatic series for 
 fermionic $\Sigma (k)$ obtained with a
 diffusive bosonic propagator again can be separated into two different subsets
 of terms: one set of terms now scales as powers of $\log \lambda$ instead of
 powers of  $\lambda$, 
 and the reduction of the expansion parameter is a direct consequence of the
 softness of bosons compared to fermions. As for phonons, the series in 
 $\log \lambda$ gives rise to vertex corrections and to the renormalization 
 of the quasiparticle residue. There are also terms that 
 form series in $\lambda$. As with phonons, these terms come from 
 boson-induced interactions between electrons and 
 their own zero-sound modes. That 
 these series hold in powers of the same $\lambda$ as the
  perturbation series with a bare boson propagator can be
   easily understood as at low frequencies, 
 the interaction between fermions and their zero-sound modes is 
 mediated by a static boson, and hence is unsensitive to any transmutation 
 of the bosonic dynamics.   
 \item
This separation of terms into perturbative series of $\lambda$ and
$\log \lambda$ allows an approximation similar to that made for
phonons to be made here.  In the magnetic case one
 neglects terms $O(\log \lambda)$ compared to terms of order $\lambda$. 
 This is not as good of a approximation as the neglect of
  $\lambda v_s/v_F$ terms 
 for phonons as $\log \lambda$  is also
 large when $\lambda$ is large. However, the numerical prefactors 
 for $\log \lambda$ series turn out to be small (a vertex correction is only 
 $(1/8) \log \lambda$), and
 in practice the neglect of logarithmical terms is well justified for all
 physically reasonable $\lambda$ ($\lambda \sim 1-2$ at optimal
 doping). In addition, from
 a purely theoretical standpoint the $\log \lambda$ terms can be 
  made parametrically small by intoducing a large number of fermionic 
 flavors $M$ (a vertex correction is then  $(1/8 M) \log \lambda$).
 Furthermore, a one-loop RG analysis of the logarithmical terms shows that
 they give rise to fractional exponents, but do not change the physics, and, 
 in particular, do not affect the pairing problem.
 \item
 As with phonons, the series in $\lambda$ yields a $\Sigma (k, \omega_m)$ that
 is predominantly dependant on frequency. More specifically, near $k-$points
 on the Fermi
surface connected by the antiferromagnetic wave vector 
${\bf Q}$ (hot spots), $\Sigma (k, \omega)$ depends on $\omega$, 
 but not on $\epsilon_k$. This momentum independence 
 is crucial for the computation of the spin polarization operator: 
 for $\Sigma (k, \omega) = \Sigma (\omega)$, the density of states is 
 flat, and $\Pi_{\omega_m}$ turns out to be independent of $\Sigma (\omega)$
 and is the same in the normal state as it would be for free fermions:  
\begin{equation}
\Pi (\omega_m) =  \frac{\omega_m}{\omega_{sf}} = 
 4 \lambda^2 \frac{\omega_m}{\bar \omega}
\end{equation}
Here we have introduced the notation
$\omega_{sf} = {\bar \omega}/(4 \lambda^2)$. 
This $\omega_{sf}$ scales as $\xi^{-2}$ and vanishes at the 
magnetic transition. 
\item   
Away from a hot spot, this independence from $\epsilon_k$ 
(i.e., on momentum perpendicular to the Fermi surface) prevails,
 but $\Sigma (k, \omega_m)$ still depends on the momentum
  {\it along} the Fermi surface. 
At $T=0$, the self-energy takes the form 
\begin{equation}
\Sigma (k, \omega) = \lambda (k)  
\frac{2 \omega}{1 + \sqrt{1 - 
i \frac{|\omega|}{\omega_{sf} (k)}}},                          \label{setot2}
\end{equation}
where 
\begin{equation}
\lambda (k) = \lambda/(1 + ({\tilde k} \xi)^2)^{1/2}, \,\,\,\,
\omega_{sf} (k) = \omega_{sf} (1 +  ({\tilde k} \xi)^2).        \label{kdep}
\end{equation} 
and ${\tilde k}$ is the component of 
${\bf k} - {\bf k}_{hs}$ along the Fermi surface.
This $k$ dependence {\it cannot} be neglected at the lowest frequencies
 as near the transition  as ${\tilde k}$ appears in a combination with $\xi$.
However, for $\omega \gg \omega_{sf} (k)$, the k-dependence disappears:
 $\Sigma (k, \omega) \approx (i \omega {\bar \omega})^{1/2}$ (we used 
the fact that $2 \lambda (k) (\omega_{sf} (k))^{1/2} = {\bar \omega}$). 
Alternatively speaking, at $\omega > \omega_{sf} (k)$, the whole Fermi
 surface acts as one big hot spot. In this range, the Eliashberg theory becomes
 applicable for all momenta.
\item
We see that whether or not the $k$ dependence of the self-energy 
can be neglected depends on what the 
 relevant $\omega$ and ${\tilde k}$ are. 
 For the pairing problem, a detailed analysis shows that 
 typical frequencies are of order ${\bar \omega}$, and typical 
 ${\tilde k}$ are of order ${\bar \omega}/v_F$. Then 
 typical $\omega_{sf} (k)$ are of order ${\bar \omega}$, i.e., 
 the momentum dependence along the Fermi surface introduces corrections 
 $O(1)$. These corrections have been checked in \cite{acf} and found to be
 nonessential from the perspective of the basic physics.
  Note also that the theory 
 assumes that ${\bar \omega} < E_F$, otherwise the linearization 
 of the dispersion near 
 the Fermi surface would not work. This in turn implies 
 that the pairing is confined to fermions in the near vicinity of a hot spot. 
\item 
In the phonon case, the momentum integration in the expressions for 
the Free energy, fermionic $\Sigma$ and anomalous vertex $\Phi$ can be
factorized and performed exactly.  Such momentum
related  corrections are always small to the 
 extent of $\lambda v_s/v_F$.  In the magnetic case, the corrections resulting
  from 
 an analogous procedure are always smaller than $1$, but whether or not they
 are small parametrically depends on the frequency. For frequencies 
 relevant to the pairing, the corections to the factorization are again $O(1)$.
\end{enumerate}

We see from the above considerations
 that at strong coupling, $\lambda \geq 1$, 
the softness of fermions compared to bosons gives rise to an effective
 Migdal theorem, i.e., the vertex corrections are smaller than $\Sigma$ 
 which in turn predominantly depends on frequency.  Contrary
 to the phonon case, there is
 no single parameter governing the validity of the Eliashberg
  approximation. There are logarithmically divergent corrections, 
 but they 
 do not affect the physics of the pairing, at least in the
 one-loop approximation.  There are also physically irrelevant 
$O(1)$ corrections stemming from the momentum dependence of 
the fermionic self-energy and the pairing vertex 
 along the Fermi surface. 
 An Eliashberg-type theory is valid when both  corrections are neglected.   
As we stated previously,
 this is quite reasonable from a physical perspective, and we now proceed
 under the assumption that the momentum dependence of $\Sigma$ and $\Phi$ 
 can be fully neglected. In the case of
 $\Phi$, this implies that we approximate the $d_{x^2-y^2}$  
 pairing vertex (and, hence the gap 
 $\Delta = \Phi \omega/(\omega + \Sigma (\omega))$) by its value
 at a hot spot, taking into account the fact that the $d-$wave
 symmetry of $\Phi$ implies that it has a different sign between hot
  spots separated by $Q$. 

\subsection{Thermodynamic potential at equilibrium}

\begin{figure}[tbp]
\epsfxsize= 3.2in
\begin{center}
\leavevmode
\epsffile{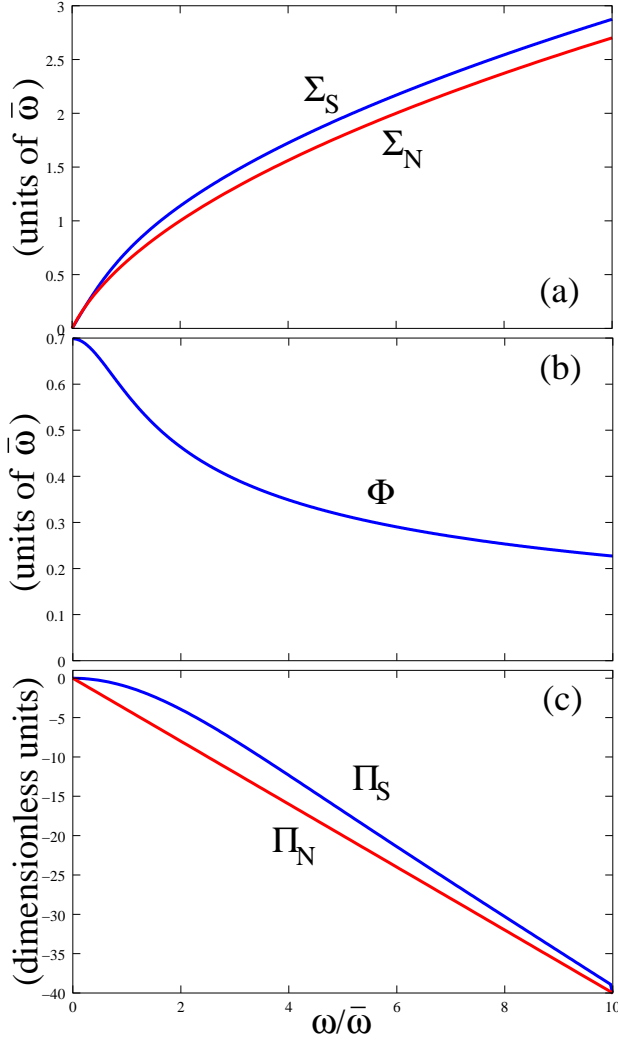}
\end{center}
\vspace{-0.75cm}
\caption{Matsubara frequency solutions at $\lambda=1$ for $\Sigma$ (a),
$\Phi$ (b) and $\Pi$ (c) in the spin fermion model (eqs \ref{eliash1})
for both the normal and superconducting states.  Note that $\Sigma$
is a strong function of frequency and may not be neglected. 
The apparent non-convergence of $\Sigma$ at high frequencies
is spurious and is discussed in the text. Further
note that $\Pi$ changes appreciably between the normal
and superconducting states.  This change {\it must} be
taken into account when calculating the condensation energy.}
\label{afmat1}
\end{figure}

We now proceed by calculating the thermodynamic potential at its
equlibrium value. 
As in the phonon case, the condition that  
 $\Omega$ is stationary 
 with respect to variations of $\Sigma$, $\Phi$ and $\Pi$ gives
\begin{eqnarray}
\Sigma(k) &=&  3 iT g^2 \chi(Q) \sum_{k'} G(k')\chi(k-k') \nonumber\\
\Phi(k) &=&  - 3 i T g^2 \chi(Q) \sum_{k'}  F(k')\chi(k-k') \nonumber\\
\Pi(k) &=& -2 T g^2\chi(Q) \sum_k  
[G(k)G(k-k') + F(k)F(k-k') ]
\label{2ndordmag}
\end{eqnarray}
Under the assumption of the momentum independence of 
$\Sigma$, $\Phi$ and $\Pi$ 
 the real and anomalous Greens functions have the form
\begin{eqnarray}
G_{\omega _{m}} ({\bf k}) &=&-\frac{\epsilon _{\mathbf{k}}+i
\widetilde{\Sigma}_{\omega_m}}{\epsilon _{\mathbf{k}}^{2}+\widetilde{\Sigma}
_{\omega_m}^{2}+\Phi^2_{\omega_m}},  \nonumber \\
F_{\omega _{m}} ({\bf k}) &=&i\frac{\Phi_{\omega _{m}}}{
\epsilon _{\mathbf{k}}^{2}+\widetilde{\Sigma}_{\omega_m}^{2}+
\Phi^{2}_{\omega_m}} 
\end{eqnarray}
and the $d_{x^2-y^2}$-pairing implies $F_\omega (\mathbf{k} +\mathbf{Q})
= - F_{\omega} (\mathbf{k})$. 
Furthermore, as we discussed, the 
 momentum integration in Eqs. \ref{2ndordmag} can be factorized and 
performed exactly. This yields
\begin{eqnarray}
\Sigma_{\omega _{m}} &=&  \lambda \pi T\sum_{n}
~\frac{\widetilde{\Sigma}_{\omega_n}}{\sqrt{\widetilde{
\Sigma}^2_{\omega_n}+\Phi^2_{\omega_n}}} 
~\frac{1}{(1 -\Pi_{\omega_{m-n}})^{1/2}},\nonumber\\ 
\Phi_{\omega _{m}} &=& \lambda \pi T\sum_{n}
\frac{\Phi_{\omega_n}}{
\sqrt{\tilde{\Sigma }^2_{\omega_n}+\Phi^2_{\omega_n}}}~
\frac{1}{(1 -\Pi_{\omega_{m-n}})^{1/2}}, 
\nonumber \\ 
\Pi _{\omega_m}(Q) &=&
  \frac{4 \lambda^2}{\bar {\omega}}  \pi T \sum_{n}[-1
 \nonumber\\ &+& 
~\frac{\widetilde{\Sigma}_{\omega_n} \widetilde{\Sigma}_{\omega_{n+m}} 
 + \Phi_{\omega_n} \Phi_{\omega_{n+m}} 
}{\sqrt{\widetilde{
\Sigma}^2_{\omega_n}+\Phi^2_{\omega_n}} \sqrt{\widetilde{
\Sigma}^2_{\omega_{n+m}}+\Phi^2_{\omega_{n+m}}}}].
\label{eliash1}
\end{eqnarray} 
We emphasize again that Eqs. (\ref{eliash1}) contain only two inputs: 
the overall energy scale $\bar \omega$ that is set 
by the spin-fermion interaction,
and the dimensionless spin-fermion 
coupling $\lambda \propto \xi$ that diverges as 
the system approaches the antiferromagnetic instability.
We also recall that the energy scale $\bar \omega$   is the 
 ultimate upper  cutoff for the strong coupling behavior 
 ($\Sigma_{\omega_m} < \omega_m$ for $\omega_m > {\bar \omega}_m$), while 
 dimensionless $\lambda$ can be represented as 
 the ratio $(2\lambda)^2 = {\bar \omega}/\omega_{sf}$ 
of ${\bar \omega}$
 and another typical scale $\omega_{sf}$ that sets the 
 upper boundary of the Fermi-liquid behavior in the  normal state.
We illustrate the form of $\Sigma_{\omega_m}$, $\Phi_{\omega_m}$ and
$\Pi_{\omega_m}$ in figure \ref{afmat1} for both the normal and superconducting
state.  

Substituting equations \ref{2ndordmag} into \ref{spinom} we obtain
 the equlibrium  thermodynamic potential in a 
magnetically mediated superconductor:
\begin{eqnarray}
\Omega &=& -T \sum_{m} \int \frac{d^2k}{(2\pi)^2} 
\lbrace  \log[ \epsilon^2_{\bf k} + \tilde\Sigma^2_{\omega_m} + \Phi^2_{\omega_m}] 
- i\Sigma_{\omega_m} G_{\omega_m} (k)
+ i\Phi_{\omega_m} F_{\omega_m} (k) \rbrace \nonumber\\
&+& \frac{3}{2} T \sum_m \int \frac{d^2k}{(2\pi)^2} 
\lbrace \log[\frac{\chi(Q,0)}{\chi(q,\omega_m)}] +
\Pi_{\omega_m} \frac{\chi(q,\omega_m)}{\chi(Q,0)} \rbrace 
\label{gfe}
\end{eqnarray}
  The difference between the above
equation and equation \ref{phononsoln1} (apart from the extra
factor of $3$) is that we have retained the bosonic part of
the free energy as it is by no means small.
In light of this, it is convenient to represent $\Omega$ as the the sum
of two parts $\Omega_{el}$ comprising the "electronic"
contributions and $\Omega_{spin}$ comprising the "magnetic"
part. Accordingly, 
 \begin{eqnarray}
 E_c = E_{c,el} + E_{c,spin} 
\end{eqnarray}
where
\begin{eqnarray}
\Omega_{el} &=&
-T \sum_{m} \int \frac{d^2k}{(2\pi)^2} 
\lbrace  \log[ \epsilon^2_{\bf k} + \tilde\Sigma^2_{\omega_m} + \Phi^2_{\omega_m}] 
- i\Sigma_{\omega_m} G_{\omega_m} (k)
+ i\Phi_{\omega_m} F_{\omega_m} (k) \nonumber\\
\Omega_{spin} &=& 
\frac{3}{2} T \sum_m \int \frac{d^2k}{(2\pi)^2} 
\lbrace \log[\frac{\chi(Q,0)}{\chi(q,\omega_m)}] +
\Pi_{\omega_m} \frac{\chi(q,\omega_m)}{\chi(Q,0)} \rbrace 
\label{cond1}
\end{eqnarray} 
The electronic term $E_{c,el} = \Omega_{el}^{S} - \Omega_{el}^{N}$ 
accounts explicitly for the appearance of the anomalous pairing vertex 
$\Phi_{\omega_n}$, {\it and} for the feedback changes to the 
fermionic self-energy.   This term by itself leads to the Wada result
for the condensation energy.  
 The term $E_{c,spin} = \Omega_{spin}^{sc} - \Omega_{spin}^{n}$
accounts for changes to the spin propagator via the
 changes to the spin polarization operator $\Pi_{\omega_m}$.
Together these two expressions account the feedback effects
between fermions and bosons in a strong coupling theory.
We point out that the distinction between $\Omega_{el}$ and $\Omega_{spin}$ is 
quite artificial, as the two are intimately connected by mutual feedback.
It is the {\it sum} of the two which is physically relevant, and the
two parts of $E_c$ may not be considered separately unless, as in the
phonon case, one of them is negligable.

As $\Sigma$ and $\Phi$ and $\Pi$ depend only on $\omega$,
the momentum integration in Eqs.\ref{cond1} can be 
performed explicitly and yields
\begin{eqnarray} 
E_{c,el} &=& - N_f \pi T \sum_m (\sqrt{\widetilde{
\Sigma}^2_{S,\omega_m}+\Phi^2_{\omega_m}} - |{\widetilde \Sigma}_{N,\omega_m}|
 \nonumber\\
 &+& |\omega_m| \frac{|{\widetilde \Sigma}_{S,\omega_m}| - \sqrt{\widetilde{
\Sigma}^2_{S,\omega_m}+\Phi^2_{\omega_m}}}{\sqrt{\widetilde{
\Sigma}^2_{S,\omega_m}+\Phi^2_{\omega_m}}})\\
\label{kin2el}
E_{c,spin} &=& -\frac{3T}{8\pi\xi^2} \sum_m \Pi_{S,\omega_m} -
\Pi_{N,\omega_m} + \log\frac{1-\Pi_{S,\omega_m}}{1-\Pi_{N,\omega_m}}
\label{sfeliash}
\end{eqnarray}
The first term $E_{c,el}$ is the Wada result.  
The second term $E_{c,spin}$ is new. Note that its expansion in  
 $\Pi_S - \Pi_N$ begins with the quadratic term. 
This is the obvious consequence of the fact that the 
Free energy is stationary with respect to a variation of $\Pi$.

\subsection{A cancellation of divergencies}

At a first glance, the electronic part of the condensation energy
 is qualitatively the same as the phonon result. This turns out,  
however, not to be the case as $E_{c,el}$ in fact contains a 
divergent piece which is cancelled out by the divergence in $E_{c,spin}$.
Indeed, consider the high frequency part of $E_{c,el}$. At high frequencies,
 ${\tilde \Sigma}_{\omega_m}$ dominates over $\Phi_{\omega_m}$, and the
 electronic part of the condensation energy reduces to 
\begin{equation}
\Omega_{el} = - 
N_f \pi T \sum_m |\Sigma_{S,\omega_m}| - |\Sigma_{N,\omega_m}| + ...
\label{26}
\end{equation} 
where dots stand for other terms that are
 all finite, as one can easily demonstrate.
Examine next the equation for $\Sigma$.
By making the substitution $\Delta_{\omega_m} = 
\Phi_{\omega_m}\omega_m/\tilde\Sigma_{\omega_m}$
we may write $\Sigma_{\omega_m}$ in the following form.
\begin{eqnarray}
\Sigma_{\omega _{m}} &=&  \pi \lambda T\sum_{n}
~\frac{\omega_n}{\sqrt{
\omega_n^2+\Delta^2_n}} 
~\frac{1}{(1 -\Pi_{\omega_m - \omega_n})^{1/2}}
\label{27}
\end{eqnarray}
Were the bosonic spectrum unchanged between the 
superconducting and normal states ($\Pi_N = \Pi_S$)
then we could expand equation (\ref{27}) 
in powers of $\Delta_n$, and would find that at large frequencies
\begin{equation}
\Sigma_{S,\omega_m} - \Sigma_{N,\omega_m} \propto 
 \frac{\Delta^2_{\omega_m}}{\omega_m}
\end{equation}
Since the gap $\Delta_{\omega_m}$ is expected on physical 
grounds to vanish at the highest frequencies (and 
 computations indeed confirm this), the frequency 
 integral in (\ref{26}) converges, i.e., the electronic 
 part of the condensation energy would be finite.
 The situation is very different
when changes in $\Pi$ are taken into account.
Although at high frequencies $\Pi_{S,\omega_n}$ indeed converges to
 $\Pi_{N,\omega_n}$, the two expressions are different at frequencies 
comparable to typical $\Delta_{\omega_m}$. 
Since for arbitrary large $\omega_m$ in equation (\ref{27}), there is
 a range of running $\omega_n$ where $\Pi_S$ and $\Pi_N$ differ,
 $\Sigma_S$ and $\Sigma_N$ do
not converge at high frequencies: 
 $\Sigma_{S,\omega_n}$ remains larger than 
$\Sigma_{N,\omega_n}$ by a constant.
We illustrate this behavior in
figure \ref{afmat1} (a).

This non-convergence of $\Sigma_N$ and $\Sigma_S$ seems at first glance
to imply an infinite result for the condensation energy!
It turns out that this apparent
infinity is compensated for by the {\it spin} part of the condensation
energy. As written in equation (\ref{sfeliash}) the spin condensation
energy
looks quite convergent {\it if} we use $\Pi_{N,\omega_m} \propto \omega_m$.
However, the expression for the spin polarization operator is formally
 ultraviolet divergent, and extra care has to be taken in evaluating
 the difference between $\Pi_{N,\omega_m}$ and $\Pi_{S,\omega_m}$.
 
In what follows we explicitly re-express the divergent contribution
 in $E_{c,el}$ in terms of the spin polarization operator, and show that
 when we take the divergent piece from $E_{c,el}$ and add it to
 $E_{c,spin}$, the dangerous $\Pi_{N,\omega_m} -\Pi_{S,\omega_m}$ term 
in $E_{c,spin}$ is cancelled out, and the remaining terms 
are all convergent, and in evaluating them we can safely use the
 regularization in which 
the
 ultraviolent divergent piece in $\Pi_{N,\omega_m}$ is absent, and 
 $\Pi_{N,\omega_m} \propto \omega_m$. In practice, this reqularization 
 amounts to evaluating the integral over $\epsilon_k$ first, and the
  frequency integral later.

\begin{figure}[tbp]
\epsfxsize= 3.2in
\begin{center}
\leavevmode
\epsffile{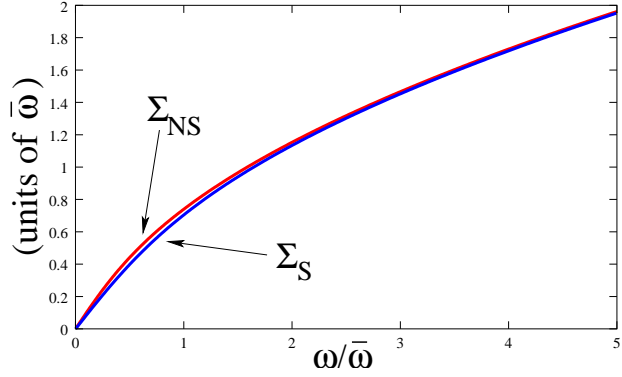}
\end{center}
\vspace{-0.75cm}
\caption{Spin-fermion solutions for $\Sigma_S$ and $\Sigma_{NS}$ 
as defined by equation \ref{sigmaNS} in Matsubara
frequencies.  Note that by replacing
$\Pi_N$ by $\Pi_S$ in the normal state expression for $\Sigma$, 
we have obtained a convergent expression for the electronic part
of the condensation energy as discussed in the text.}
\label{afmat2}
\end{figure}

In order to accurately single out the divergent piece in
 $E_{c,el}$ and relate it to the spin polarization operator, 
 we use a trick originally suggested by Bardeen and Stephen and 
 define
a mixed self energy $\Sigma_{NS}$ ($NS$ stands for normal-superconducting).
This is the normal state Eliashberg equation for $\Sigma$ but with
the {\it superconducting} polarization bubble.
\begin{eqnarray}
i\Sigma_{NS,\omega _{m}} &=& -\alpha^2 \pi T \sum_n \int \frac{d^2q}{(2\pi)^2}
\chi_{S,\omega_n}(q)G_{N,\omega_{n+m}}(k+q) \nonumber\\
&=&  \pi \lambda T\sum_{n}
sgn (\omega_n) 
~\frac{1}{(1 -\Pi_{S,\omega_{m-n}})^{1/2}}
\label{sigmaNS}
\end{eqnarray}
We plot $\Sigma_{NS}$ given by the above equation along with
$\Sigma_S$ in figure \ref{afmat2} and show that they converge
at high frequencies.
We then add and subtract $\Sigma_{NS}$ from $E_{c,el}$.
\begin{eqnarray}
E_{c,el} &=& - N_f \pi T \sum_m (\sqrt{\widetilde{
\Sigma}^2_{S,\omega_m}+\Phi^2_{\omega_m}} - |{\widetilde \Sigma}_{NS,\omega_m}|
 \nonumber\\
 &+& |\omega_m| \frac{|{\widetilde \Sigma}_{S,\omega_m}| - \sqrt{\widetilde{
\Sigma}^2_{S,\omega_m}+\Phi^2_{\omega_m}}}{\sqrt{\widetilde{
\Sigma}^2_{S,\omega_m}+\Phi^2_{\omega_m}}}) \nonumber\\
&+& \lbrace |\tilde\Sigma_{NS,\omega_m}| - |\tilde\Sigma_{N,\omega_m}| \rbrace
\end{eqnarray}
One can easily make sure that 
$E_{c,el}$ now consists of a convergent piece
plus the divergent $|\tilde\Sigma_{NS}| - |\tilde\Sigma_{N}|$.  
The above is actually $|\Sigma_{NS}| - |\Sigma_N|$ as 
the $\omega_m$'s cancel.

We now explicitly express this divergent piece in terms of the
 spin polarization operator. To accomplish this, we recall that
the term $|\tilde\Sigma_{N}|$ in $E_{c,el}$ arose from
the integration over momentum of the $\Sigma G$ term in Eq. \ref{cond1}.
Writing this for both the normal and normal-superconducting self
energies, we have:
\begin{eqnarray}
N_f \pi T \sum_m |\Sigma_{N,NS,\omega_m}| = T\sum_m \int \frac{d^2k}{(2\pi)^2}
i\Sigma_{N,NS,\omega_m} G_{N,\omega_m} 
\end{eqnarray}
By using the expressions for $\Sigma$ and $\Pi$ 
 the above may be written as a term
in $E_{c,spin}$ as follows:
\begin{eqnarray}
&+& T\sum_m \int \frac{d^2k}{(2\pi)^2}
i\Sigma_{N,NS,\omega_m} G_{N,\omega_m} \nonumber\\
&=& T \sum_m \int \frac{d^2k}{(2\pi)^2} \lbrace -3g^2 \pi T \sum_n 
\int \frac{d^2q}{(2\pi)^2} \chi_{N,NS,\omega_n}(q) 
G_{N,\omega_{n+m}}(k+q) \rbrace G_{N,\omega_n}(k) \nonumber\\
&=& -\frac{3}{2} T \sum_n \int \frac{d^2q}{(2\pi)^2} 
\frac{\chi_{N,NS,\omega_n}(q)}{\chi(0)}
\lbrace 2 g^2 \chi(0) \pi T\sum_m 
\int \frac{d^2k}{(2\pi)^2}  
G_{N,\omega_{n+m}}(k+q) G_{N,\omega_n}(k) \rbrace \nonumber\\
&=& -\frac{3}{2\chi_0} T \sum_n \int \frac{d^2q}{(2\pi)^2} 
\chi_{N,NS,\omega_n}(q) \Pi_{N,\omega_n}
\end{eqnarray}
Performing the integration over q we find that
\begin{equation}
-\pi T N_f \sum_m  \left( |\Sigma_{NS,\omega_m}| - 
|\Sigma_{N,\omega_m}|\right)  
= - T \frac{3}{8\pi \xi^2} \sum_m {\Pi}_{N, \omega_m} 
\log{\frac{1  -{\Pi}_{S,\omega_m}}
{1 - {\Pi}_{N, \omega_m}}}
\label{matNfNs}
\end{equation}
We now move the divergent piece from $\Omega_{el}$ to $\Omega_{spin}$ and 
write the condensation energy as $E_c = \delta\tilde\Omega_{el} 
+ \delta\tilde\Omega_{spin}$ with
\begin{eqnarray} 
\delta \tilde\Omega_{el} &=& - N_f \pi T \sum_m (\sqrt{\widetilde{
\Sigma}^2_{S,\omega_m}+\Phi^2_{\omega_m}} - |{\widetilde \Sigma}_{NS,\omega_m}|
 \nonumber\\
 &+& |\omega_m| \frac{|{\widetilde \Sigma}_{S,\omega_m}| - \sqrt{\widetilde{
\Sigma}^2_{S,\omega_m}+\Phi^2_{\omega_m}}}{\sqrt{\widetilde{
\Sigma}^2_{S,\omega_m}+\Phi^2_{\omega_m}}})\\
\label{kin2el_1}
\delta\tilde\Omega_{spin} &=& -\frac{3T}{8\pi\xi^2} \sum_m \Pi_{S,\omega_m} -
\Pi_{N,\omega_m} + (1+\Pi_{N,\omega_n})\log
\frac{1-\Pi_{S,\omega_m}}{1-\Pi_{N,\omega_m}}
\label{convergmat}
\end{eqnarray}
The electronic part is now fully convergent. For the spin part, 
one can easily check that at large frequencies, when $\Pi_{N,\omega_m}$ 
and $\Pi_{S,\omega_m}$ are both large, the expansion of the logarithm
 in $E_{c,spin}$ cancels
 the dangerous 
$\Pi_{S,\omega_m} - \Pi_{N,\omega_m}$ term. The remaining terms are all 
 ultraviolet convergent, i.e., are insensitive to the regularization
  procedure used to evaluate $\Pi_{N,\omega_m}$. 
This implies that the condensation energy is actually free from 
divergencies, as it indeed should be based on  physical reasoning.

\subsection{The relation between $\delta {\tilde \Omega}_{el}$ and 
$\delta {\tilde \Omega}_{spin}$.}

 At this point, the electronic and spin contributions to the 
condensation energy seem to be rather different as the electronic part 
 contains $N_f$, while the spin part doesn't. However, 
 $\delta {\tilde \Omega}_{el}$ and $\delta {\tilde \Omega}_{spin}$ are 
 in fact of the same order 
 as we  now demonstate. Indeed, as we already said, typical frequencies 
 for the pairing are of order ${\bar \omega}$, and at these frequencies 
\begin{equation}
\Pi_{S,\omega_n} \sim \Pi_{N, \omega_n} \sim \frac{\bar \omega}{\omega_{sf}}
\end{equation}
Similarly, 
\begin{equation}
\Sigma_{S,\omega_n} \sim \Sigma_{N, \omega_n} \sim {\bar \omega}  
\end{equation}
Then using equation \ref{convergmat} for $\delta\tilde\Omega_{spin}$:
\begin{equation}
\delta {\tilde \Omega}_{spin} \sim \frac{{\bar \omega}^2}
{\omega_{sf} \xi^2} \sim \frac{\bar \omega^3}{v^2_F}
\end{equation}
where the last step uses the definition of $\omega_{sf}$ given
previously.
At the same time
\begin{equation}
\delta {\tilde \Omega}_{el} \sim N_f {\bar \omega}^2
\label{aa1}
\end{equation}
The Fermionic density of states $N_f$  is a product of $1/v_F$ 
(the leftover of the integration over $\epsilon_k$, and a typical 
${\tilde k}$ along the Fermi surface). As typical ${\tilde k} 
\sim {\bar \omega}/v_F$, 
\begin{equation}
N_f \sim \frac{\bar \omega}{v^2_F}
\end{equation}
Substituting this result into (\ref{aa1}), we find
 \begin{equation}
\delta {\tilde \Omega}_{el} \sim \frac {{\bar \omega}^3}{v^2_F},
\label{aa2}
\end{equation}
i.e., $\delta {\tilde \Omega}_{el}$ and $\delta {\tilde \Omega}_{spin}$ are
 indeed of the same order. 

In the above discussion  $N_f$ appears as an 
 {\it extra} parameter in 
$\delta {\tilde \Omega}_{el}$. This is
 because in the calculations we neglected the 
momentum dependence along the Fermi surface (the actual momentum integral 
 over $d {\tilde k}$ is replaced by a typical ${\tilde k}$). 
 If this momentum dependence was included (i.e., by using
  the self-energy from  
Eq. (\ref{setot2}) with $k-$dependent $\lambda (k)$ and $\omega_{sf} (k)$, 
 then the electronic part would be free from uncertainties. 
Unfortunately, this computation also requires the knowledge of the 
$k-$dependence of $\Phi (k)$ along the Fermi surface, which is 
technically difficult to obtain.  
In constrast, 
$\delta {\tilde \Omega}_{spin}$ is the result of a full two-dimensional 
integration over bosonic momenta, and the result for 
$\delta {\tilde \Omega}_{spin}$ is free from uncertainties.

Fortunately, it turns out that within the (approximate) 
computational scheme that we are using, $N_f$ and 
$1/(\omega_{sf} \xi^2)$ can be  related. Their relation follows  from 
 Eq. (\ref{matNfNs}), as both the fermionic self-energy and the spin 
 polarization operator are fully expressed in terms of ${\bar \omega}$ and 
 $\lambda$. By evaluating the constant pieces in 
 $\Sigma_{NS, \omega_m} - \Sigma_{N, \omega_m}$ and in
  $\Pi_{S,\omega_m} - \Pi_{N, \omega_m}$ at high frequencies
 and comparing the two sides of Eq. (\ref{matNfNs}), one
  can express $N_f$ in terms of $1/(\omega_{sf} \xi^2)$.
   Once this is done, there is no further uncertainty in the
    condensation energy - it is given by the universal function of $\lambda$ times 
${\bar \omega}^2/(\omega_{sf} \xi^2) \propto {\bar \omega}^3/v^2_F$.

At the risk of belaborating this point, we note
 that the fact that $E_{c,el}$ and $E_{c,spin}$ 
 are of the same order implies that one
  cannot replace Wada's expression for 
 for electronic part of the condensation energy by a Bardeen-Steven formula. 
Indeed, the ability to do this in a phonon superconductors was based on 
 the integral relation between $\Sigma_{N, \omega}$ and $\Sigma_{S,\omega}$, 
Eq. \ref{relation}. For our case, the r.h.s. of this relation is
 finite and is given by  
\begin{equation}
K= \frac {3}{8 \omega_{sf} \xi^2} \int_0^{\infty} d \omega_m \log 
{\frac{\omega_{sf} + \omega_m}{\omega_{sf} + \omega_m f_{\omega_m}}}
 S(\omega_m) 
\end{equation}
 where $f_{\omega_m} = \Pi_{s,\omega_m}/\Pi_{n,\omega_m}$, and 
\begin{equation}
S(\omega_m) =  \int_0^{\infty} d \omega_n \frac{\Phi^2_{\omega_n}}
{\sqrt{\widetilde{
\Sigma}^2_{s,\omega_n}+\Phi^2_{\omega_n}} +{\widetilde \Sigma}_{s,\omega_n}
 \sqrt{\widetilde{
\Sigma}^2_{s,\omega_n}+\Phi^2_{\omega_n}}} + 
\int_0^{\omega_m} d \omega_n ~\frac{{\widetilde \Sigma}_{s,\omega_n}}{
\sqrt{\widetilde{
\Sigma}^2_{s,\omega_n}+\Phi^2_{\omega_n}}}
\end{equation}
A simple order of magnitude estimate shows  that $K$ is of the same order as 
$\Sigma_{s,\omega_m}$, i.e., for magnetic superconductors, 
the transformation from Wada's to Bardeen and Steven's  formula for
 the electronic condensation energy introduces a correction
  of the same order as $E_{c,el}$.

One final remark.  The electronic
 and spin parts of $E_c$ are only of the same order of magnitude
 as long as ${\bar \omega} < E_F$. When the effective coupling
 exceeds $E_F$, typical ${\tilde k} = O(1)$, i.e., the whole Fermi surface is
 involved in the pairing. In this limit, the spin-fermion
 calculations are not controlable. Estimates show, however,
  that typical frequencies for
 the pairing now scale as 
$ \omega_{sf} \xi^2 \sim v^2_F/{\bar \omega} \sim J$ where $J$ 
is the exchange  integral for the corresponding Heisenberg model.
 (Recall that ${\bar \omega} \sim g^2 \chi(Q)/\xi^2$ where  the
  RPA approximation, $g$ is equivalent to Hubbard $U$. In the 
  same approximation, near a magnetic transitiion $\chi(Q) \sim \xi^2/U$, 
  i.e., ${\bar \omega} \sim U$.)  Estimating $\delta {\tilde \Omega}_{spin}$
   at typical frequencies, we indeed find 
$\delta {\tilde \Omega}_{spin} \sim J$ in agreement with the result by 
Scalapino and White~\cite{scal_white}. The same reasoning yields 
$\delta {\tilde \Omega}_{el} \sim N_f~J^2 \sim J^2/v_F \ll J$. We see 
 therefore that at very large couplings, the spin part of the condensation 
 energy  clearly prevails over the electronic part, i.e., the 
 condensation energy comes entirely from the spin part. This 
 again agrees with Scalapino and White~\cite{scal_white}.

\section{The computations}

In this section we present our results for the electronic
 and spin contributions to the condensation energy for various $\lambda$. 
In practice, we found it advantageous to perform the calculations of 
$\delta {\tilde \Omega}_{spin}$ and $\delta {\tilde \Omega}_{el}$ 
in real frequencies rather than in Matsubara frequencies. The main 
reason for this was simply
that we had previously evaluated  $\Sigma (\omega)$, 
$\Phi (\omega)$ and
 $\Pi (\omega)$ at real frequencies and various couplings and could use these
 results in the present computations. A more subtle reason is that in 
 retarded formalism,  the problem of divergencies in $E_{c,el}$ and 
 $E_{c,spin}$  can be avoided in a straightforward
 manner. (see below)

\subsection{Condensation energy in real frequencies}

We first derive the expression for the condensation energy in real frequencies, in terms of retarded $\Sigma (\omega)$, $\Phi (\omega)$ and $\Pi (\omega)$. The Matsubara equations for $E_{c,el}$ and $E_{c,spin}$
given in equation \ref{cond1} have the
following form.
\begin{eqnarray} 
E_{c,el} &=& -\pi T \sum_m f(i\omega_m) \nonumber\\
E_{c,spin} &=& - \pi T \sum_n g(i\omega_n)
\end{eqnarray}
where the $E_{c,el}$ has a sum over {\it fermionic}
frequencies and $E_{c,spin}$ has a sum over {\it bosonic}
frequencies.  The retarded form of these equations, assuming no
branch cuts except on the real axis are
\begin{eqnarray}
E_{c,el} &=& -\int_0^{\infty} f"_{ret}(\omega) \tanh(\frac{\omega}{2T})
d\omega \nonumber\\
E_{c,spin} &=& -\int_0^{\infty} g"_{ret}(\omega) \coth(\frac{\omega}{2T})
d\omega
\end{eqnarray}
where  $f"$ is the imaginary part of f ($f'$ is the real part) and
$g$ is similar.
It remains to analytically continue $f(\omega_m)$ and $g(\omega_n)$ to
the real axis.  With the Matsubara definitions used in section 2, the analytic
continuations are as follows.
\begin{eqnarray}
\Sigma(\omega_m) \rightarrow -i\Sigma_{ret}(\omega) \nonumber\\
\Phi(\omega_m) \rightarrow \Phi_{ret}(\omega) \nonumber\\
\Pi(\omega_m) \rightarrow \Pi_{ret}(\omega)
\end{eqnarray}
The retarded formulas for the condensation energy are then
\begin{eqnarray}
E_{c,el} &=& - N_f \int_0^\infty \lbrace
[\beta + Re \tilde\Sigma_N(\omega) ] \nonumber\\
&+& \omega[1- \frac{Im \Sigma_S(\omega)~ \alpha - Re \Sigma_S (\omega)~ \beta}
{|\alpha|^2 + |\beta|^2}] \rbrace \tanh\frac{\omega}{2T} d\omega \nonumber\\
E_{c,spin} &=& - \frac{3}{8\pi^2 \xi^2} \int_0^\infty
\lbrace Im \Pi_S - Im \Pi_N 
\nonumber\\
&+& Im \log\frac{1-\Pi_S(\omega)}{1-\Pi_N(\omega})
\rbrace \coth\frac{\omega}{2T} d\omega
\end{eqnarray}
Where $\sqrt{\Phi^2(\omega) - \Sigma_S^2(\omega)} = \alpha+ i\beta$.
We point out that extreme care must be taken with these equations
in order to get the correct sign of the imaginary parts of both the square
root and the logarithm.  

\begin{figure}[tpb]
\epsfxsize= 3.2in
\begin{center}
\leavevmode
\epsffile{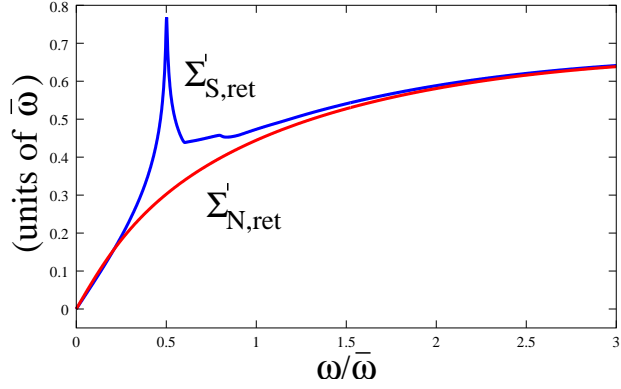}
\end{center}
\vspace{-0.75cm}
\caption{The real part of the retarded self-energy 
for both the normal and superconducting states.  
Observe that $\Sigma'_{S,ret}$
and $\Sigma'_{N,ret}$ converge at high frequencies in contrast to 
the constant offset in Matsubara frequencies.  The constant offset
goes into the imaginary part of $\Sigma_{ret}$ which does not affect
$E_c$ in the retarded formulism.}
\vspace{-0.5cm}
\label{ret}
\end{figure}

We first point out that there is no divergent term in $E_{c,el}$.
Indeed, in the Matsubara formalism, the divergent term 
 comes from the fact that at high frequencies, $\Sigma_S(\omega_n)$
and $\Sigma_N(\omega_n)$ are separated by a constant. Since we defined  
$\Sigma$ with an extra $i$, this constant is {\it imaginary}.   
On the other hand, the the first two terms in the retarded formula for 
$E_{c,el}$
 at high frequencies where
$\Phi \rightarrow 0$  can be written as
\begin{eqnarray}
Im  \sqrt{-\Sigma^2_S(\omega)} + Re \Sigma_N(\omega)
=  Re \Sigma_N(\omega) -Re \Sigma_S(\omega)
\end{eqnarray}
This follows from the fact that $\Sigma_S(\omega) = \Sigma'(\omega)
 + i|\Sigma"(\omega)|$
and the branch cut is on the negative real axis.
We see that $E_{c,el}$ only depends on the difference of 
$Re \Sigma$ between the normal and superconducting states, and the integral
of this difference is
 fully convergent. We illustrate this in figure \ref{ret}
Analogous reasoning also shows that $E_{c,spin}$ is also free from
 divergencies. 

Indeed, the absence of divergencies in the retarded formalism is just 
the consequence of using the Kramers-Kronig transform which misses the 
divergent pieces 
 in $E_{c,spin}$ and $E_{c,el}$. However, since
  we already demonstrated that the full $\delta \Omega$ is free 
  from divergencies, we
 can safely use the Kramers-Kronig transformation separately for 
 $E_{c,spin}$ and $E_{c,el}$. 

The fact that no divergence exists for the retarded formulas also 
 allows us to relate the prefactors
in front of $E_{c,el}$ and $E_{c,spin}$
in a straightforward manner.  
In real frequencies, Eq. \ref{matNfNs} takes the form 
\begin{eqnarray}
&&-N_f \int_0^{\infty} \lbrace   Re \Sigma_{NS}(\omega) - 
Re \Sigma_{N} (\omega) \rbrace \tanh \frac{\omega}{2T} d\omega \nonumber\\
=&&  \frac{3}{8\pi^2 \xi^2} \int_0^\infty  \lbrace Re \Pi_N(\omega)
  Im  \log \frac{1-\Pi_S(\omega)}
{1-\Pi_N(\omega)} \nonumber\\
 &&+ Im \Pi_N(\omega) Re \log \frac{1-\Pi_S(\omega)}
{1-\Pi_N(\omega)} \rbrace \coth \frac{\omega}{2T} d\omega
\label{retNfNs}
\end{eqnarray}
Since in the retarded formalism, $ \int_{-\infty}^{\infty}
Re \Sigma_{NS}(\omega)- Re \Sigma_N(\omega) d\omega$ is a 
convergent quantity, and the r.h.s. of (\ref{retNfNs}) is also convergent,
 we can  explicitly evaluate (numerically) both sides of (\ref{retNfNs}) and
 relate $N_f$ and $3/8 \pi^2 \omega_{sf} \xi^2$ which we label as spin density of states $N_s$ ($N_s = (8/3\pi^2){\bar \omega}/v^2_F$).   

\subsection{Results}

As we have already stated, we use previously obtained results for 
$\Sigma (\omega)$, $\Phi (\omega)$ and $\Pi (\omega)$. 
First, we  computed both sides of Eq, (\ref{retNfNs}) and 
evaluated the ratio $N_f/N_s$ for various $\lambda$. We found that 
 with very small variations, $N_f/N_s \approx 5.9$. 

\begin{figure}[tbp]
\epsfxsize= 3.2in
\begin{center}
\leavevmode
\epsffile{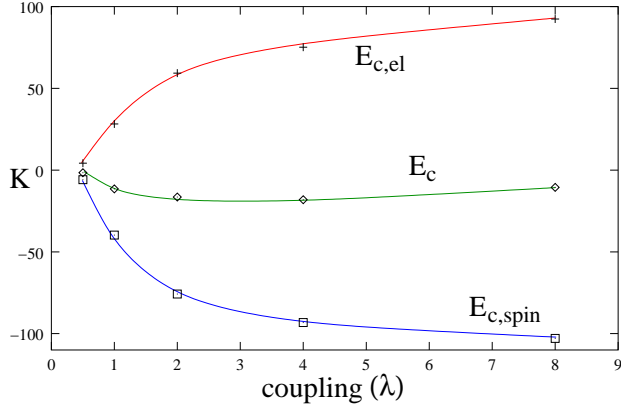}
\end{center}
\vspace{-0.75cm}
\caption{Electronic ($E_{c,el}$),  
spin ($E_{c,spin}$), and total ($E_c$) condensation energy per
unit cell at $T=0$ for various couplings
$\lambda$. The lines
are a guide for the eye.  The sum of $E_{c,el}$ and $E_{c,spin}$
produces a total condensation energy which is {\it negative}.
 We used $N_f = 1 st/eV$, and $N_s \sim 0.17 st/eV$ as explained
in the text.}
\label{fig:1}
\end{figure}

In Fig.\ref{fig:1} we present the results for the 
electronic and spin contributions to the condensation energy 
for different values of the coupling $\lambda$.
To set the overall scale, we 
adopt a commonly used estimate $N_f = 1st/eV$~\cite{abanov}.
We emphasize that
changing $N_f$ will only change the overall
scale and not the functional form of $E_c(\lambda)$.

We see from Fig. \ref{fig:1} that the total condensation energy is
negative, as it indeed should be in a superconductor, but that
this negativity is of a very different origin than in BCS theory.
In BCS theory, which corresponds to $\lambda \ll 1$, 
i.e.,  ${\bar \omega} < \omega_{sf}$, the system behaves as a
 conventional Fermi liquid.
 In this limit, the pairing potential is static, i.e., 
the spin part of $E_c$ is negligible, and  
condensation energy is entirely electronic and {\it negative}.
We see however for $\lambda \geq 1$, ie the strongly coupled
regime, the 
the electronic contribution  to
the condensation energy is {\it positive} and quite large.
 From the figure
it appears that the electronic contribution changes sign at 
 $\lambda \sim 0.4$. $E_{c,el}$
is negative below this coupling strength 
and is positive for all $\lambda \geq 1/2$ presented 
in the figure. 
Second,  for all $\lambda$ shown,
the spin part $E_{c,spin}$ 
is negative. It can be shown that  $E_{c,spin}$ continues to be
negative at $\lambda \rightarrow 0$
where $\Pi \rightarrow 0$ as $\lambda^2$. (see Eq. \ref{eliash1})
Indeed, by expanding the logarithm in Eq. \ref{retNfNs} we obtain
\begin{eqnarray}
E_{c,spin}(\lambda \rightarrow 0) &=& -N_s \frac{\omega_{sf}}{2} 
\int_0^{\infty}
Im  \lbrace \Pi_S^2 - \Pi_N^2 \rbrace \nonumber\\
&=& -N_s \omega_{sf} \int_0^{\infty}
\lbrace Re \Pi_S Im \Pi_S - Re \Pi_N Im \Pi_N \rbrace < 0
\end{eqnarray}
The above equation is negative, as in the retarded formalism,
$Re \Pi_N=0$ and both $Re \Pi_S \le 0$ and $Im \Pi_S<0$, and scale
as $\lambda^4$ as $\lambda\rightarrow 0$. 

We also see from the figure that at large $\lambda$, 
both the spin and the electronic 
parts of the condensation energy nearly saturate:
to a large positive value for $E_{c,el}$ and a large 
negative value for $E_{c,spin}$. 
The total condensation energy $E_c$ is negative
and much smaller than either $E_{c,el}$ or
$E_{c,spin}$ due to a substantial cancellation 
between these two components of $E_c$. 
Although this cancelation seems quite delicate,
it is actually robust since $E_{c,el}$ and $E_{c,spin}$
are intimately linked via mutual feedback, and can not
be considered separately.  it is the sum of the two which has physical meaning. 
 Any estimate of the total condensation energy based
merely upon either the electronic or spin part will give a highly erroneous
result.

\begin{figure}[tpb]
\epsfxsize= 3.2in
\begin{center}
\leavevmode
\epsffile{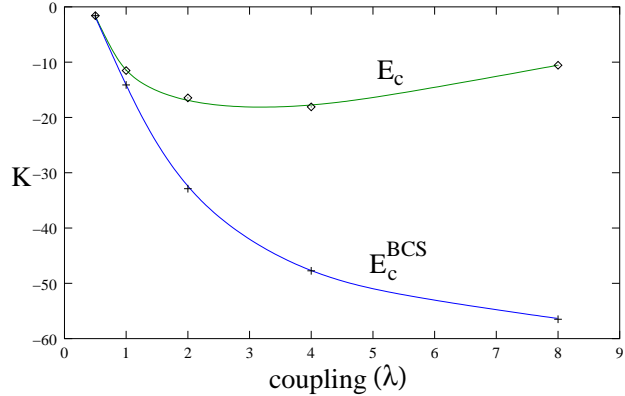}
\end{center}
\vspace{-0.75cm}
\caption{Total condensation energy $E_c$ compared
with the BCS result $E_c^{BCS}$ at $T=0$ for various
couplings $\lambda$. We used $N_s \sim 0.17 st/eV$ and $N_f/N_s \sim 5.9$ as 
 explained in the text. Observe that the  BCS condensation energy 
 monotonically increases as the coupling gets larger, while 
 the actual 
condensation energy  flattens at 
 $\lambda \sim 2$ and slightly 
decreases at large couplings.}
\label{fig:2}
\end{figure}

We now consider the functional dependence of $E_c$ on $\lambda$.
In figure \ref{fig:2}, we see that  the condensation energy flattens 
at $\lambda \sim 2$, and decreases
 at large couplings despite the fact that the pairing gap 
 increases monotonically with $\lambda$~\cite{acf}.
 This behavior is 
very counterintuitive from a BCS perspective, 
 where the condensation energy scales with $\Delta^2$.  
It clearly indicates that for $\lambda \geq 1$, the physics
 is qualitatively different from BCS theory. 
To emphasize this strong deviation from 
BCS theory we plot in Fig.\ref{fig:2} 
the strong coupling result of $E_c$ along with
 the BCS condensation energy $-N_f\Delta^2/2$ using the same 
$\Delta$ and $N_f$.  We clearly see that for $\lambda \geq 1$, corresponding
 to optimally doped and  underdoped cuprates, BCS theory yields
  qualitatively incorrect results for $E_c$.

 Our results are in line with earlier work which demonstrated that  
 for $\lambda \geq 1$,  the  pairing  predominantly involves 
 fermions located in the non-Fermi liquid frequency range. For these fermions,
 retardation effects not included in BCS theory become dominant.
 Such retardation effects take place between the
 ``upper'' - ${\bar \omega}$  and ``lower'' $\omega_{sf}$ scales of  
 spin-fermion theory.   As $\bar\omega = 4 \lambda^2 \omega_{sf}$
 this ratio grows quickly, and already $\bar\omega/\omega_{sf} = 4$
 at $\lambda=1$.  This explains why the deviations from BCS 
 behavior are already strong at this coupling.  
Understanding in detail the strong coupling physics behind the 
decrease in $E_c$ is currently the subject of a separate
 study~\cite{artem} and a complete theory of this phenomenon does
 not exist at the moment.
 Most likely, however, this decrease is a reflection of the fact that 
as $\lambda$ increases, the actual attraction between fermions goes down, 
 retardation of the spin-mediated interaction becomes the major factor, and  
the pairing process 
increasingly involves incoherent (diffusive) fermions and 
on-shell bosons.   As the exchange of 
on-shell bosons is an energy conserving process, it  
 can not lead to a gain in $E_c$.
Such behavior is very counterintuitive from a BCS perspective, 
 where the pairing emerges due to an exchange of virtual, 
 off-shell bosons, and the condensation energy scales with $\Delta^2$. 

One final comment.  Although magnetically mediated superconductors
are often compared to dirty superconductors, we point out that
 at $T=0$, the physics of the two is already qualitatively different.
Analogies between the two are often made due
to the fact that {\it thermal} spin fluctuations 
 scatter at finite momentum transfer but zero energy transfer and  
act in the same way as non-magnetic impurities~\cite{acf}. 
However, in a dirty superconductor with non-magnetic impurities, 
the condensation energy retains its BCS form despite of the fact that  
the superfluid stiffness is renormalized down~\cite{dirty}. 
Obviously, this is not what we found. 

\section{Kinetic Energy}

As we stated in the introduction, 
several groups have argued~\cite{mike1,hirsh}
 that the condensation energy is 
 driven by a gain in the kinetic energy which at strong
  coupling is negative (in contrast to BCS theory) 
 because of a strong ``undressing'' of fermions which 
  bear a greater resemblance to free particles in the superconducting state 
than they do in the normal state.  

In this section we consider, within our model, the 
 change in the kinetic energy when the system enters 
the superconducting state.
The conventionally defined kinetic-energy for an interacting fermionic system 
 is  
\begin{equation}
E_{kin} = 2  T\sum_m  \int \frac{d^2 k }{(2\pi)^2}~ \epsilon_k~ G_{\omega_m} (k) 
\label{kin}
\end{equation}
where $G_{\omega_m} (k)$ is the full fermionic Green's function 
that contains the self-energy. Integrating over momentum and subtracting the
normal state result from $E_{kin}$ in a superconductor 
we obtain
\begin{equation}
\delta E_{kin} = 2N_f \pi T \sum_m \sqrt{\widetilde{
\Sigma}^2_{S,\omega_m}+\Phi^2_{\omega_m}} - |{\widetilde \Sigma}_{N,\omega_m}|
\label{simpkin}
\end{equation}
  In the BCS limit, $\lambda \ll 1$, 
$\Phi_{\omega_n} = \Delta$, 
$\Sigma =0$, ${\widetilde \Sigma}_{\omega_n} = \omega_n$ and 
\begin{eqnarray}
\delta E_{kin}^{BCS}=  2N_f \pi T \sum_m \sqrt{
\omega_m^2+\Delta^2} - |{\omega_m}|
\label{kin-bcs} 
\end{eqnarray}
which is obviously positive and furthermore depends logarithmically 
 on  the upper limit of the frequency integration, 
which is $\omega_{sf}$ in our case
 (we recall that in the BCS limit, $\omega_{sf} >> \bar\omega$).
  At $T=0$, we have 
\begin{equation}
\delta E_{kin}^{BCS}=  N_f \Delta^2 \log{\frac{\omega_{max}}{\Delta}}
\label{kin-bcs_1}
\end{equation}
In the same BCS limit, the potential part of the condensation energy 
$\delta E_{pot}^{BCS}$ is also logarithmically divergent, 
and to a logarithmical accuracy cancels out $\delta E_{kin}^{BCS}$.
 The subleading terms do not cancel and yield 
 $E_c^{BCS}= - N_f \Delta^2/2$.

We now consider finite $\lambda$.
As before, we perform the computations in  real frequencies.  
The analytic continuation
of equation \ref{simpkin} gives
\begin{eqnarray}
\delta E_{kin} = +2 N_f \int_0^\infty \lbrace Im  
\sqrt{\Phi^2(\omega) - \widetilde\Sigma_S^2(\omega)} + Re 
\widetilde\Sigma_N(\omega)
\rbrace \tanh \frac{\omega}{2T} d\omega
\end{eqnarray}

\begin{figure}[tbp]
\epsfxsize= 3.2in
\begin{center}
\leavevmode
\epsffile{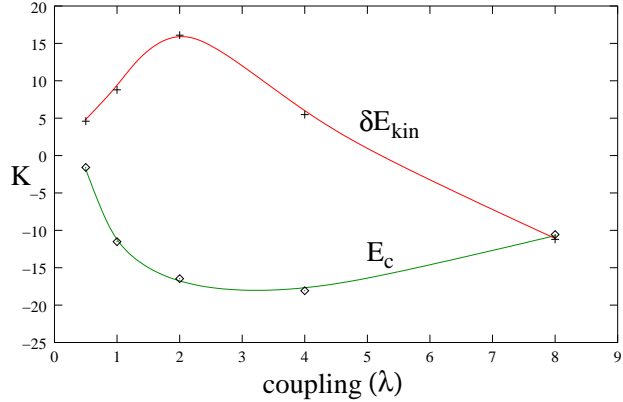}
\end{center}
\vspace{-0.75cm}
\caption{Kinetic energy $\delta E_{kin}$ compared with total condensation energy
$E_c$ at $T=0$ for various couplings $\lambda$.
The parameters are the same as in Fig.  2.  The kinetic energy change 
is positive at low couplings, but negative at high coupling.}

\label{fig:3}
\end{figure}

The result of this calculation
 at finite $\lambda$ are presented in Fig~\ref{fig:3}.
At low couplings the kinetic energy is positive, as one naively
expects.  At larger $\lambda$, however, the kinetic energy passes through a
maximum at $\lambda \sim 2$ and then becomes negative at 
large $\lambda$. 

As we already mentioned in the Introduction, the sign of $E_c$
 depends on the interplay between two competing effects: the 
 effect of particle-hole mixing that increases $E_{kin}$, and  the 
change in the self-energy 
due to  the ``undressing'' of fermions   that lowers  $E_{kin}$. 
At weak coupling, the particle-hole mixing obviously dominates. 
The sign change between small and large $\lambda$  implies
 that at strong coupling the situation is reversed, and  the 
 lowering of $E_{kin}$ via the change in the self-energy 
due to  the ``undressing'' of fermions   
overcomes the effect of particle-hole mixing 
   This behaviour is very similar to that
obtained by Norman et. al. \cite{mike1}.
 As the first term in $E_{c,el}$ is equal to $-\delta E_{kin}/2$
 (see Eqs. (\ref{kin2el} and \ref{simpkin}),
 one can indeed argue that 
 the condensation energy at large couplings  is at least partly 
 driven by the lowering of the kinetic energy.  However, 
 a comparison of Figs. \ref{fig:1} and \ref{fig:3} shows that 
 this is just another way to interpret strong coupling effects that 
 affect {\it both} the fermionic and bosonic propagators via mutual feedback.
  
A simple explanation  of why this is so is the following. In the  
 superconducting state, the spin decay into fermions is forbidden
  at energies smaller than $2\Delta$.
This  {\it simultaneously} gives rise to two effects. First, 
 the spin propagator develops 
 the excitonic (resonance) peak at $\omega_{res} < 2\Delta$. 
  The energy released by the creation of an exciton results in a 
  gain in the magnetic part of the condensation energy. 
 Secondly, the fermions cannot  decay 
 until their frequency exceeds $\Delta + \omega_{res}$ (this is  the
 magnetic analog of the Holstein effect).
 The elimination of fermionic scattering at low
  frequencies implies that  the fermionic self-energy $\Sigma (\omega)$ 
  in the superconducting state is reduced compared to 
that in the normal state. 
 This lowers the kinetic energy. Obvously, the two effects 
(the gain in the magnetic part and the lowering of the kinetic energy) 
 come from the same physics.

\section{Conclusions}

Our goal in this paper was to emphasize the importance of taking
all contributions to the condensation energy into account when
considering a strongly coupled  superconductor.  Specifically,
we considered the case of 
$d_{x^2-y^2}$ pairing mediated by the exchange of near-critical 
overdamped antiferromagnetic spin fluctuations. 
 We demonstrated that 
 although Eliashberg theory is valid for a strongly coupled
 magnetic superconductor, the reason for its validity is
  qualitatively different from that for phonon superconductors as the
  spin velocity and Fermi velocity are of the same order. Due to this fact, 
approximations appropriate for phonon superconductors are generally
{\it not} valid in magnetic superconductors.  Specifically, we
 demonstrated that
the assumption that the  bosonic polarization bubble can be neglected, 
 which was rigorously justified by 
 Bardeen and Stephen for phonon superconductors, 
breaks down for magnetically-mediated superconductors and makes the Wada 
and Bardeen-Stephen formalisms invalid.

We obtained the full expression for the condensation energy within 
the spin-fermion model and showed that
 the spin and electronic parts of the condensation energy are of
  the same order ${\bar \omega}^3/v^2_F$ and both depend only on
   the dimensionless coupling 
$\lambda$. The BCS behavior is restored at $\lambda \ll 1$.
 Even at moderate couplings, the condensation energy
 is highly non-BCS. The electronic
 contribution to the condensation 
energy is {\it positive}, while the spin part  
is negative and larger in magnitude than the electronic part which makes 
the full $E_c$ negative. 
  As in the BCS limit  the
electronic condensation energy is negative and equal to $-N_f \Delta^2/2$,
this implies that the electronic condensation energy 
changes sign at a rather small $\lambda$. 
We found that  at large $\lambda$, both the spin and the electronic 
 parts of the condensation energy nearly saturate.
As a result,  the full condensation energy flattens 
at $\lambda \sim 2$, and decreases
 at large couplings despite the fact that the pairing gap 
 increases monotonically with $\lambda$~\cite{acf}. This behavior is 
very counterintuitive from a BCS perspective, 
 where the condensation energy scales with $\Delta^2$ and is also 
 inconsistent with the behavior of $E_c$ in dirty superconductors.
This behavior results from the fact 
that there is a substantial cancellation between the 
 spin and electronic parts of the condensation energy and the total $E_c$
 is thus substantially smaller than either the spin or electronic parts.

We argued that the reduction of $E_c$ at large coupling is likely the 
result of the fact that at strong coupling the pairing is predominantly
 due to an energy
 conserving  exchange of on-shell (real) bosons as opposed to the BCS 
 theory in which the pairing is caused by the energy non-conserving exchange of 
virtual bosons.  The reduction of $E_c$ at strong coupling also indicates that 
 in this limit coherent superconductivity becomes fragile, and a large 
$\Delta$ only indicates that the system needs a finite energy to destroy 
 spin singlets.

Finally, we computed the kinetic energy and found that at strong coupling 
 it is negative which indicates
 that at high couplings  the change in the self-energy 
due to  the ``undressing'' of fermions, which lowers $E_{kin}$    
overcomes the effect of particle-hole mixing which tends to increase $E_{kin}$.
This behavior has no analog for phonon superconductors.
We argued that a negative $E_{kin}$ is fully consistent with a positive 
$E_{c,el}$, and that, in principle,  it is correct to argue that 
 the condensation energy at large couplings  is at least partly 
 driven by the lowering of the kinetic energy.  However, 
  our results show that the lowering of $E_c$ may be thought of
  as being due to either the 
the lowering of the kinetic energy, {\it or} the 
interplay between the lowering of $\Delta \Omega_{spin}$ and 
the increase of $E_{c,el}$. 
 Both explainations are valid interpretations of the strong coupling effects which 
 affect the fermionic and bosonic propagators via mutual feedback.
  
By taking into account {\it all} contributions to the self energy,
and taking $N_f \sim 1 st/eV$ as typical for near optimally doped cuprates, 
we obtained a small value for $E_c$ of $\sim 15 K$ at optimal doping,
 (which in our model corresponds to $\lambda \sim 1.5-2$~\cite{abanov_advances},
 larger $\lambda$ describe underdoped cuprates).
 This is rather remarkable as all typical energies in the problem are much 
higher, i.e ${\bar \omega} \sim 2.5-3 \times 10^3 K$~\cite{abanov_advances}. 
This small value of $E_c$ is partly 
due to small prefactors, but is also the result of substantial
cancellation between the spin and electronic contributions to $E_c$.
Note also that $N_f \sim 1 st/eV$ that we were using is
 fully  consistent with  ${\bar \omega} \sim 2.5-3 \times 10^3 K$.
Indeed, using Fermi surface averaged $v_F \sim 0.6 meV$ ~\protect\cite{acf}   
 we obtain $N_s = (8/3\pi^2){\bar \omega}/v^2_F \sim 0.15-0.19 st/eV$.
Using then $N_f/N_s \sim 5.9$ obtained in the paper, we find $N_f \sim 1 st/eV$, i.e., precisely the same value as we used  

Our $E_c \sim 15 K$ is in  good agreement with experiment.  
Loram et al~\cite{loram} extracted  
$E_c \approx 0.12 k_BT_c \sim 10 K$ 
from the jump of the  specific heat at $T_c$. 
The change of the functional form of $E_c$ around $\lambda =2$ is also 
 consistent with the experimental fact that $E_c$ changes its behavior
  from BCS-like to non-BCS  around optimal doping.  
A decrease of $E_c$ at strong couplings (i.e., for underdoped cuprates) 
is also consistent with what Loram {\it et al} 
found in the specific heat experiments in 
the underdoped regime~\cite{loram}. We caution, however, that the
 relation between $E_c$ and the amount of the jump in the specific 
 heat at $T_c$, from which the experimental $E_c$ was extracted may
  be more complex than in the BCS theory which was used to
   extract $E_c$ from the data. This analysis is clearly called for.

We thank Ar. Abanov, B. Altshuler, 
G. Blumberg, J. Loram,  D. van-der-Marel, K. Pepin,  D. Pines, 
 E. Yuzbashyan,
 and especially M. Norman 
 for useful discussions. This research was supported by
  NSF DMR-9979749 (A. Ch)
and  by DR Project 200153,  and the Department
 of Energy, under contract W-7405-ENG-36. (R.H.) 

\section{Appendix A - A running coupling constant approach}

The condensation energy may also be computed using a 
 general formula for the ground state energy of the interacting 
electron system~\cite{lutt_ward,agd,mahan}:
\begin{equation}
E - E_0 = - i \int_0^\lambda~ \frac{d \lambda_1}{\lambda_1} 
\int \frac{d^2 k d \omega}{(2\pi)^3}~~G_{\omega_m} (k)  \Sigma^*_{\omega_m} 
\label{n1}
\end{equation}
where $E_0$ is the ground state energy of free electrons,
 and the Green's function and the self-energy are evaluated for 
 the running coupling constant $\lambda_1$. 
 This formula is valid {\it both} in the normal and superconducting state,
The ``effective'' self-energy 
 $ \Sigma^*_{\omega_m}$ is related to $G_{\omega_m} (k)$ in the
  same way as a conventional self-energy, i.e., as
$G^{-1}_{\omega_m} (k) = i (\omega_m + \Sigma^*_{\omega_m}) - \epsilon_k$.   
In the normal state, 
 $\Sigma^*_{n,\omega_m} = \Sigma_{n,\omega_m}$, where 
$\Sigma$ is a conventional self-energy, while  in 
 the superconducting state, 
\begin{equation}
\Sigma^*_{s,\omega_m}= \Sigma_{s,\omega_m} + 
\frac{i \Phi^2_{\omega_m}}{\epsilon_k + i {\widetilde \Sigma}_{s,\omega_m}}
\label{star}
\end{equation}

Eq. (\ref{n1}) is particularly suitable for the strong coupling 
computations in the normal state. Here we can use 
\begin{equation}
\int \frac{d^2k}{4\pi^2} G_{\omega_m} (k) = - i \pi N_f sign \omega_m
\end{equation}
Subsituting this result into (\ref{n1}) we reduce $E - E_0$ to a single frequency integral. 
Using the $T=0$ normal state result for the spin-fermion model 
$\Sigma_{n,\omega_m} = 2\lambda 
\omega_m/(1 + (1 + |\omega_m|/\omega_{sf})^{1/2})$, and 
and introducing the sharp upper cutoff for the low-energy
 theory at  $\omega_{max} \sim E_F$, we obtain 
\begin{equation}
E-E_0 = - N_f (\omega^3_{max}{\bar \omega})^{1/2} 
\int_0^{2 \lambda(\frac{\omega_{max}}{\bar \omega})^{1/2}}
~\frac{d x}{(\sqrt{1 +x^2} +1)^2}
\end{equation}
This expression is convenient for the analysis of the 
variation of the ground state energy with $\lambda$.

 The condensation energy, i.e., the energy difference between normal and 
 superconducting state, is given by
\begin{equation}
E_c = -i 
\int_0^\lambda~ \frac{d \lambda_1}{\lambda_1} 
\int \frac{d^2 k d \omega}{(2\pi)^3}~~(G_{s,\omega_m} (k) 
 \Sigma^*_{s\omega_m} - G_{n,\omega_m} (k)  \Sigma^*_{n\omega_m} )
\label{n2}
\end{equation}
Using $G \Sigma^* = -i (1 - G^{-1}_0 G)$,
we can rewrite (\ref{n2}) as 
\begin{equation}
E_c = N_0 \int \frac{d \epsilon_k d \omega_m}{2\pi}~
 \int_0^{\lambda} \frac{d \lambda_1}{\lambda_1}~(G_{s,\omega_m} (k) -
 G_{n,\omega_m} (k)) (i \omega_m - \epsilon_k)
\label{n8}
\end{equation}
Performing the momentum integration in the Green's functions, we obtain
\begin{equation}
E_c = - N_f  
 \int_0^{\lambda} \frac{d \lambda_1}{\lambda_1}
 \int_0^{\infty} d \omega_m
  \left(\sqrt{\widetilde{
\Sigma}^2_{s,\omega_m}+\Phi^2_{\omega_m}} - {\widetilde \Sigma}_{n,\omega_m}
 - \omega_m \frac{{\widetilde \Sigma}_{s,\omega_m} - \sqrt{\widetilde{
\Sigma}^2_{s,\omega_m}+\Phi^2_{\omega_m}}}{\sqrt{\widetilde{
\Sigma}^2_{s,\omega_m}+\Phi^2_{\omega_m}}}\right)
\label{n3}
\end{equation} 
This looks very similar to the Wada result 
( Eq. \ref{kin2el}) but note the minus sign in front of the $\omega_m$.
This extra
minus sign results in a negative
total $E_c$ for all couplings.

The condensation energy, Eqn. (\ref{n8}) can also be formally divided 
 into kinetic and potential energy terms, but this division is subjective for 
 interacting systems, and we will not discuss this.

\subsection {BCS limit}
In the BCS limit, $E_c$ reduces to
\begin{equation}
E_c = - N_f \int_0^{\lambda} \frac{d \lambda_1}{\lambda_1}~
 \Delta^2_{\lambda_1} \int_0^{\infty} \frac{d \omega_m}
 {\sqrt{\Delta^2_{\lambda_1} + \omega^2_m}}
\label{n31}
\end{equation}
where $\Delta$ is the gap value for the running coupling $\lambda_1$.
Using the BCS relation between $\Delta$ and the coupling constant
\begin{equation}
1 = \lambda_1 N_f \int_0^{\infty} \frac{d \omega_m}{\sqrt{\Delta^2 + \omega^2_m}}
\label{n4}
\end{equation}
one can rewrite (\ref{n31}) as
\begin{equation}
E_c = -  \int_0^{\lambda} \frac{d \lambda_1}{\lambda^2_1}~\Delta^2_{\lambda_1}
\label{n5}
\end{equation}
From (\ref{n4}), 
\begin{equation}
\Delta_{\lambda_1} = \Delta_{\lambda} e^{\frac{1}{N_f \lambda} - \frac{1}{N_f \lambda_1}}
\label{n6}
\end{equation}
Integrating over $\lambda_1$ we obtain
\begin{equation}
E_c = - \frac{1}{2}~ N_f \Delta^2   
\label{n7}
\end{equation}
This is indeed the same result as we obtained using the Eliashberg formula.

Note in passing that our previous assertion that the 
 separation of $E_c$ into a kinetic and potential energy is subjective
 is true even in the BCS limit, as the interaction is the source of 
the pairing. Indeed, earlier we computed $\delta {E}_{kin}$ in the Luttinger-Ward formalism and found  that in the BCS limit, the 
kinetic energy scales as $\Delta^2 \log {\omega_{max}/\Delta}$ (see Eq. (\ref{kin-bcs_1})). In the running coupling constant formalism, the  
 kinetic energy difference 
 $\delta {\tilde E}_{kin}$ extracted from (\ref{n8}) in the BCS limit reduces to
\begin{equation}   
\delta {\tilde E}^{BCS}_{kin} = 2 N_0 \int_0^{\lambda} \frac{d \lambda_1}{\lambda_1}
 \int_0^{\infty} \frac{\Delta^2_{\lambda_1} d \omega_m}{\omega_m + 
\sqrt{\Delta^2_{\lambda_1} + \omega^2_m}}
\label{n9}
\end{equation}
Using (\ref{n4}) and performing the computations with the 
logarithmical accuracy we find 
$\delta {\tilde E}^{BCS}_{kin} = N_0 \Delta^2/2$, i.e., contrary to 
 Eq. (\ref{kin-bcs_1}) the newly defined kinetic energy does not depend 
 logarithmically on the upper limit of frequency integration. Similarly,
$\delta {\tilde E}^{BCS}_{pot} = -  N_0 \Delta^2$ such that the sum of
 the two yields
 the correct total condensation energy, Eq. (\ref{n7}). This once again 
 demonstrates that  only total $E_c$ is a physically meaningful quantity.

\subsection{$\lambda =1$} 

We numerically 
computed $E_c$ for $\lambda=1$ by calcuating the integrand
of eqn \ref{n3} for $\lambda=0$, $0.5$ and $1$,
 numerically fitting these three
data points and then integrating the resultant function over the coupling constant.
This resulted in a condensation energy of $E_c=-15.8 K$ compared
to the $-11.5 K$ calcuated via the Eliashberg approach.  This agreement
is very impressive given the highly approximate nature of the calculation.
Unfortunately, the running coupling constant formalism is very computationally
intensive, hence our motiviation 
for using the Eliashberg approach in this paper.

\section{Appendix B - A peculiarity of the BCS approximation}
\label{app_bcs}

In this Appendix, we elaborate on our earlier discussions
of the result for the 
condensation energy in the BCS limit.  In BCS theory,  the pairing
 problem may be described by an effective quadratic Hamiltonian
\begin{equation}
H = H_0 + 
\sum_{k,\alpha} \epsilon_k ~a^\dagger_{k,\alpha} a_{k, \alpha} +
\frac{\Delta}{2} g_{\alpha,\beta} (a^\dagger_{k, \alpha} 
a^\dagger_{-k,\beta} + a_{k, \beta} a_{-k,\alpha})
\label{a1}
\end{equation} 
where $g_{\alpha \beta}$ is the antisymmetric matrix~\cite{agd}. 
The condensation energy can then be straightforwardly
 obtained by simply averaging both the normal and anomalous terms in (\ref{a1}) Expressing the average products of the pairs of operators in terms
 of frequency integrals of normal and anomalous Freen functions,
  and taking care to avoid
 double counting of the anomalous term, we obtain the zero temperature result
\begin{equation}
E_c = - \frac{N_f \Delta^2}{2} I
\label{a2}
\end{equation}
where
\begin{equation}
I = \int \frac{d \epsilon d \omega}{\pi} 
~\frac{\omega^2 - \epsilon^2}{(\epsilon^2 + \omega^2)
 (\epsilon^2 + \omega^2 + \Delta^2)}
\label{a3}
\end{equation}
We know that the correct result is $I=1$. However, a naive integration
 treating
 both $\omega_m$ and $\epsilon$ in  (\ref{a3}) on an equal footing, 
 results in a vanishing integral. This vanishing is surely artificial,
  as the 2D integral over $d \epsilon d \omega$ 
 is logarithmically divergent and therefore the result does depend on  
the order of limits of the integration. This situation is a 
typical example of  an anomaly.
 
The correct way is to perform the integration over $\omega$ first as in the 
 Hamiltonian approach to the pairing, the interaction is independent of
 frequency, and hence there is no cutoff in the frequency integration. 
 The integration over energy, on the other hand, is obtained by simplifying the original integral over momenta and is  
is only valid as long as the 
 density of states is a constant which is true up to an 
upper cutoff $\Lambda \sim \epsilon_F$. 
 Integrating first 
 over $\omega_m$ in infinite limits, and then integrating 
over $\epsilon_k$ we obtain
\begin{equation}
I = 2 \int_0^{\Lambda/\Delta} \left(\sqrt{x^2 +1} - 2 x + 
\frac{x^2}{\sqrt{x^2 +1}} \right)
\label{a4}
\end{equation}
One can easily make sure that the integral converges at large $x$ and 
therefore does not depend on the upper limit as long as 
$\Delta \ll \Lambda$. Setting the upper limit to infinity 
 we easily obtain 
 $I =1$ as it indeed should be.

It is also instructive to reproduce the correct result 
by integrating over $\epsilon_k$ first. 
A formal integration in the infinite limit 
 yields an incorrect  $I = -1$,
 However, integrating in 
(\ref{a3}) first 
over $\epsilon_k$ between $-\Lambda$ and $\Lambda$, and then 
over $\omega$, we obtain 
\begin{equation}
I = - \frac{2}{\pi}  \int_0^{\infty} dx \left( \frac{1 + 2 x^2}{\sqrt{x^2 +1}} 
\tan^{-1} {\frac{\Lambda^*}{\sqrt{x^2 +1}}} - 2 x \tan^{-1} 
{\frac{\Lambda^*}{x}} \right)
\label{a5}
\end{equation}
where $\Lambda^* = \Lambda/\Delta$.
If we formally set $\Lambda^* = \infty$, the integral will be convergent and 
 yield an incorrect result, $I = -1$. However, keeping $\Lambda^*$ 
 large but finite we find after changing variables  to $y = x/\Lambda^*$
\begin{equation}
I = -1 + \frac{8}{\pi}(\Lambda^*)^2 \int_0^{\infty} dy y \left( 
 \tan^{-1} \frac{1}{y} - \tan^{-1} \frac{1- (2 y^2 (\Lambda^*)^2)^{-1}}
{y} 
 \right)
\label{a6}
\end{equation}
Expanding under $\tan^{-1}$ in $1/(\Lambda^*)^2$ and evaluating the
 remaining integral we find
\begin{equation}
I = -1 + \frac{4}{\pi} \int_0^{\infty} \frac{d y}{y^2 +1} = -1 +2 =1
\label{a7}
\end{equation}
as it indeed should be.

The physical implication of this result is that in the BCS theory, 
the condensation energy can be equally viewed as coming from the energy levels 
  near the
  Fermi surface, as implied in Eq. (\ref{a4}) where the 
  integral is confined to $x = O(1)$, i.e., to $\omega \sim \Delta$,  
or as coming from very deep levels below the Fermi surface, 
as implicated in Eq. (\ref{a6}) where the integral is confined 
to $y = O(1)$, or to $\omega \sim \Lambda \sim E_F$. Such an uncertainty 
is typical for an anomaly which 
can be equivalently viewed as coming from   
either low energies or high energies~\cite{chiral}.

This peculiarity, however, is only present in the BCS limit, where the gap 
 remains finite even at the largest frequencies (this is what causes 
 logarithmical divergencies in the integrals). In the Eliashberg theory,
  the gap vanishes at infinite frequency, the integrals are convergent, 
  and the condensation energy can only be viewed as coming from the 
  levels near the Fermi surface.

\subsection{A relation between $E_c$ in Refs. \cite{mike1} and \cite{review}}

The peculiar nature of the BCS limit also explains the apparent sign difference
 between the expressions for the condensation energy in  Refs. \cite{mike1}
 and \cite{review}. In Ref.~\cite{review}, Scalapino derived the condensation energy by
averaging the interaction electron-phonon Hamiltonian:
\begin{equation}
<H> = -i  \int \frac{d^d k d \omega}{(2\pi)^{d+1}} 
(\omega + \epsilon_k) G (k, \omega) - <\sum_\nu \frac{P^2_\nu}{M}>
\label{aaa1}
\end{equation}
where the last term is twice the expectation value of the ion kinetic energy. He then used
 Chester's result~\cite{chester} for the  relation between 
the isotopic dependence of the upper critical field 
 and  the change in the ion kinetic energy between normal and
 superconducting states,  
 and found that for the  isotope exponent $\alpha =1/2$, 
the 
change in the ion kinetic energy is precisely  minus twice the
 change in the electronic propagator. As a result, 
the condensation energy turns out to be minus the difference 
between the first terms in (\ref{aaa1}) in a superconductor and the normal state:
\begin{equation}
E_c = <H_{sc}> - <H_n> = - i \int \frac{d^d k d \omega}{(2\pi)^{d+1}} 
(\omega + \epsilon_k) 
~\left(G_n (k, \omega) - G_s (k, \omega) \right)
\label{aaa2}
\end{equation}
[Note that we define $E_c$ with the opposite sign compared to Refs. \cite{mike1} 
and \cite{review} - our $E_c$ is negative, while their $E_c$ is positive].

For a nonzero self-energy
 $\Sigma = \Sigma (\omega)$, the momentum and frequency integral in 
(\ref{aaa2}) is ultraviolet convergent 
because the pairing gap $\Delta = \Delta (\omega)$ 
vanishes at $\omega \rightarrow \infty$, and  the ordering of the momentum 
 and frequency integration does not matter. 
Performing integration over frequency first, one ontains, 
after simple manipulations, Wada's formula, Eqn. (\ref{wada}). 
In the BCS limit of vanishing $\Sigma (\omega)$ and a constant 
$\Delta$, this yields $E_c = - N_0 \Delta^2/2$, as we discused earlier.

The authors of Ref~\cite{mike1},  on the other hand, assumed that the
 condensation energy is not due to phonons, and that
  $\alpha$ is nearly zero, as in the near optimally doped cuprates. They 
 argued that in this situation, the 
 condensation energy should be given  solely by the first term in (\ref{aaa1}),
 i.e., 
\begin{equation}
E_c =  -i \int \frac{d^d k d \omega}{(2\pi)^{d+1}} (\omega + \epsilon_k) 
~\left(G_s (k, \omega) - G_n (k, \omega) \right)  =  \int \frac{d^d k}{
(2\pi)^d} \int^0_{-\infty} (\omega + \epsilon_k) 
\left(A_s (k, \omega) - A_n (k, \omega) \right) 
\label{aaa3}
\end{equation} 
where $A(k, \omega) = (1/\pi) Im G (k, \omega)$ is the quasiparticle spectral function.
This expression has opposite sign compared to Eqn (\ref{aaa2}). 
Still, the authors of Ref.\cite{mike1} argued that their expression 
 also reproduces the BCS result $E_c = - N_0 \Delta^2/2$.

The consideration in the previous subsection of this appendix shows that both expressions are
 correct in the BCS limit (despite the fact that the one is minus the other!). The 
 way in which the condensation energy was obtained in ref \cite{mike1} implies 
 that the momentum and frequency integral must be physically motivated, i.e., 
 if the momentum integral over $d^d k \approx
 N_0 d \epsilon_k$  is extended to infinite limits, 
one has to integrate first over momentum and then over frequency.
On the contrary, the Wada expression for the condensation energy implies that
 the momentum integral comes first, and the frequency integration comes second.
As we already know, in the BCS limit, interchanging the order
 of the integration changes the result by a factor $-1$. This explains why the
 two apparently opposite results for $E_c$ actually yield the same condensation energy. 

A comparison of the two results shows that that in the BCS limit,
 the change in the ion kinetic energy  merely  sets the 
 proper regularization of the ultraviolet divergent integral for $E_c$.
If one regularizes the integral by restricting the momentum integration 
 to $|\epsilon_k| \leq  \Lambda \sim E_F$ 
(this is what we called the physically
 motivated regularization), the ion kinetic energy may be 
completely neglected. 
Reversing the ordering of the integration  is equivalent to imposing 
  a constraint on the frequency integral. This constraint can only 
come from the frequency dependence of the phonon propagator, i.e., 
from the kinetic energy of
 ions. Obviously  in this situation, the ion kinetic energy 
cannot be neglected.

Away from the BCS limit, the correct way to proceed in the general case 
is to use the
 Luttinger-Ward-Eliashberg expression for Free energy, Eq. 
(\ref{phononom}). This expression is valid for arbitrary $\Sigma (k, \omega)$
 and it also includes the full feedback on bosons. Note that the expression 
for $\Omega$  is more general 
 than the Eliashberg theory for phonon superconductivity as 
Eqn (\ref{phononom}) actually
 doesn't assume that $\Sigma$ depends only on frequency, 
and that one can factorize the
 momentum integration in the expressions for the self-energy and 
the pairing vertex,
 Eqs (\ref{2ndord}). The only approximation 
 in the Eliashberg formula is the neglect of vertex corrections 
(which account for higher-order terms labled by dots in (\ref{phononom})).
Still, it is of interest to understand which of the 
 two expressions for $E_c$, Eqn \ref{aaa2} or \ref{aaa3}, one should 
use for the situation where the feedback effect on  the bosonic
 propagator can, for one reason or another, be neglected. 
We argue that the answer depends on whether the self-energy 
predominantly depends on 
$k$ or on $\omega$. 

Indeed, in the BCS limit, the physical motivation for the
 ordering of the momentum and frequency integrations
 (first over frequency and then over $\epsilon_k$) was
associated with the fact that, for a constant gap, 
the ultraviolet divergence of the momentum and frequency integral for $E_c$
 was cut by that the integration over $\epsilon_k$ cannot extend to 
$|\epsilon_k| > \Lambda \sim E_F$. 
Suppose now  that the 
 Eliashberg theory is valid, i.e., $\Sigma \approx \Sigma (\omega)$, and 
$\Delta = \Delta (\omega)$. As we already said, 
this $\Delta (\omega)$ vanishes above some characteristic frequency 
$\omega_0$ which decreases when fermion-boson coupling increases. 
Once this frequency becomes smaller than $\Lambda$, the cutoff of the
 ultraviolet divergence is provided by the frequency integral, 
rather than momentum 
integral. In this situation, the physically motivated
ordering of the integrations should be integration over $\epsilon_k$ first 
(in infinite limits), and the integration over frequency afterwards. 
This ordering is implicit in Eq. \ref{aaa2}. Not surprisingly, this
 expression yields Wada formula which in turn follows from the Eliashberg 
 Free energy provided that $\Sigma = \Sigma (\omega)$. If instead we
 used Eqn. (\ref{aaa3}), we would obtain a rapid variation of $E_c$ 
 once ${\bar \omega}$ becomes smaller than $\Lambda$, and eventually 
Eqn (\ref{aaa3}) would yield the result opposite in sign to Eq (\ref{aa2}).

On the other hand, when the self-energy predominantly depends on $k$, the
 cut of the ultraviolet divergence  is still provided by the momentum integral.
 In this situation, the ionic kinetic energy can be neglected, 
and Eqn. \ref{aaa3} should be used. This can be explicitly verified by 
comparing the Luttinger-Ward-Eliashberg formula with the bosonic term dropped,
 Eqn. (\ref{aaa4}), with Eqn. (\ref{aaa3}). 
Setting  $\Sigma (k, \omega) = \Sigma (\epsilon_k)$ and
 $\Phi(k, \omega) = \Phi (\epsilon_k)$ in  (\ref{aaa4}) and  integrating 
over frequency we obtain 
\begin{eqnarray} 
E_c &=& - N_f \int_0^\infty d \epsilon 
(\sqrt{\widetilde{
\Sigma}^2_{S,\epsilon}+\Phi^2_{\epsilon}} - |{\widetilde \Sigma}_{N,\epsilon}|
 \nonumber\\
 &+& \epsilon \frac{|{\widetilde \Sigma}_{S,\epsilon}| - \sqrt{\widetilde{
\Sigma}^2_{S,\epsilon}+\Phi^2_{\epsilon}}}{\sqrt{\widetilde{
\Sigma}^2_{S,\epsilon}+\Phi^2_{\epsilon}}})
\label{wada_1}
\end{eqnarray}  
 where ${\widetilde \Sigma}_\epsilon = \epsilon + \Sigma (\epsilon)$.
This expression is the analog of the Wada formula for the condensation energy
(\ref{wada}). 
We now do the same in Eqn. (\ref{aaa3}). The frequency integration is 
 straightforward and by proper evaluation of the arguments of the logarithms
 we obtain {\it exactly the same expression as Eqn. (\ref{wada_1})}.
This proves our point that in theories where the feedback from the pairing 
on bosonic propagator can be neglected,  Eq. (\ref{aaa4}) is valid for
 any $\Sigma (k, \omega)$, while Eqs. (\ref{aaa2}) or (\ref{aaa3}) are valid 
when $\Sigma (k, \omega) \approx \Sigma (\omega)$ and
  $\Sigma (k, \omega) \approx \Sigma (k)$, respectively.

\end{document}